\documentclass[
    twocolumn, superscriptaddress, nofootinbib, amsmath, amssymb,aps, floatfix, longbibliography
]{revtex4-1}
\usepackage{xr}
\usepackage[dvipsnames,table]{xcolor}
\usepackage{graphicx,epsfig,psfrag,bm,amssymb}
\usepackage[colorlinks=true,urlcolor=blue
,anchorcolor=blue,citecolor=blue,filecolor=blue,linkcolor=blue,menucolor=blue
]{hyperref}
\usepackage{dcolumn}
\usepackage{bm}
\usepackage{soul}
\usepackage{cancel}
\usepackage{mathrsfs,amsfonts,color}
\usepackage{caption}
\usepackage{subcaption}
\usepackage{comment}
\usepackage{xspace}
\usepackage[capitalise]{cleveref}
\usepackage{ulem}
\usepackage{cancel}
\newcommand{\eV}{{\mathrm{eV}}}

\newcommand{\vecbf}[1]{{\bm{#1}}}
\newcommand{\beq}{\begin{equation}}
\newcommand{\eeq}{\end{equation}}
\newcommand{\gag}{g_{a\gamma}}
\newcommand{\Jdm}{\vecbf{J}_\mathrm{D}}
\newcommand{\mdm}{m_{\mathrm{D}}}
\newcommand{\Psig}{P_\text{sig}}

\newcommand{\tx}[1]{\ensuremath{\textnormal{#1}}}
\newcommand{\eqa}[1]{\begin{align}#1\end{align}}

\newcommand{\rp}{\right)}
\newcommand{\lp}{\left(}
\newcommand{\rb}{\right]}
\newcommand{\lb}{\left[}
\def\bit{\begin{itemize}}
\def\eit{\end{itemize}}
\def\ben{\begin{enumerate}}
\def\een{\end{enumerate}}
\newcommand{\la}[1]{\label{#1}}
\newcommand{\Eq}[1]{Eq.~\eqref{#1}}

\newcommand{\Tab}[1]{Table~\ref{#1}}

\newcommand{\pfrac}[2]{\lp\frac{#1}{#2}\rp}

\newcommand{\units}[1]{\;\tx{#1}}

\newcommand{\GeV}{\units{GeV}}

\newcommand{\cm}{\units{cm}}

\newcommand{\micron}{\mu \text{m}}

\newcommand{\be}{\begin{equation}}
\newcommand{\ee}{\end{equation}}

\DeclareMathOperator{\Max}{Max}

\let\Re\relax
\DeclareMathOperator{\Re}{Re}

\newcommand{\Order}{\mathcal{O}}

\newcommand{\bra}[1]{\langle #1 | }
\newcommand{\ket}[1]{| #1 \rangle}
\newcommand{\ip}[2]{\langle #1 | #2 \rangle}

\newcounter{questioncount}[section]

\newcommand{\ldb}{\lambda_\tx{dB}}

\DeclareMathOperator{\re}{Re}
\DeclareMathOperator{\im}{Im}

\makeatletter
\AddToHook{cmd/appendix/before}{\def\cref@section@alias{appendix}\def\cref@subsection@alias{appendix}}
\makeatother

\makeatletter
\newcommand*{\addFileDependency}[1]{
\typeout{(#1)}
\@addtofilelist{#1}
\IfFileExists{#1}{}{\typeout{No file #1.}}
}\makeatother

\begin{document}

\preprint{FERMILAB-PUB-24-0034-T}
\title{Scalable Dark Matter Searches Using Integrated Photonics}
\author{Nikita Blinov}
\affiliation{Department of Physics and Astronomy, York University, Toronto, Ontario, M3J 1P3, Canada}
\author{Christina Gao}
\affiliation{Department of Physics, Southern University of Science and Technology, Shenzhen, 518055, China}
\author{Roni Harnik}
\affiliation{Theoretical Physics Division, Fermi National Accelerator Laboratory, Batavia, IL 60510, USA}
\author{Ryan Janish}
\affiliation{Theoretical Physics Division, Fermi National Accelerator Laboratory, Batavia, IL 60510, USA}
\author{Neil Sinclair}
\affiliation{John A. Paulson School of Engineering and Applied Sciences, Harvard University, 29 Oxford St., Cambridge, MA 02138, USA}
\date\today

\begin{abstract}
Dark matter (DM) with masses of order an electronvolt or below can have a non-zero coupling to electromagnetism while being compatible 
with cosmological observations. 
In these models, the ambient DM behaves as a new classical source in Maxwell's equations, 
which can excite potentially detectable electromagnetic (EM) fields in the laboratory.
We propose a new integrated-photonics–based approach to search for dark matter candidates in the 0.1–few eV mass range.
This approach offers a wide range of wavelength-scale devices like resonators and waveguides that are readily fabricated in large quantities, enabling a scalable and novel search. 
In particular, we demonstrate that refractive index-modulated resonators, such as etched/grooved microrings, or 
patterned slabs, support EM modes with efficient coupling to DM. 
When excited by DM, these modes are read out by coupling the resonators to a waveguide that terminates on a micron-scale-sized single photon detector, such as a single pixel of a  low-noise charge-coupled device or a superconducting nanowire. 
We then estimate the sensitivity of this experimental concept in the context of axion-like particle and dark photon models of DM, 
demonstrating that nanophotonic confinement and scalability can extend dark matter sensitivity into previously unexplored parameter space.
\end{abstract}

\maketitle
\section{Introduction}
Dark matter (DM) constitutes 85\% of all matter in the universe, yet its microscopic nature remains completely unknown~\cite{Bertone:2016nfn}. 
Among the most promising candidates are ultralight particles with masses far below the electron mass—a largely unexplored regime that could revolutionize our understanding of fundamental physics. 
Theoretical frameworks including string theory and solutions to the strong CP problem naturally predict ultralight bosonic particles: axions~\cite{Abbott:1982af,Dine:1982ah,Preskill:1982cy,Arias:2012az}, axion-like particles (ALPs)~\cite{Svrcek_2006}, and dark photons (DPs)~\cite{Abel:2008ai,Nelson:2011sf,Arias:2012az}. 
Unlike conventional DM candidates, these particles would form coherent, wave-like fields oscillating at frequencies determined by their mass.  

This wave-like behavior opens entirely new detection strategies. When coupled to electromagnetic (EM) fields, such DM acts as a classical source in Maxwell's equations, potentially driving detectable EM fields in laboratory devices. However, current experimental approaches face severe limitations in the electronvolt mass range,
whereas theoretically wave-like DM masses can span many orders of magnitude including the eV scale and above~\cite{Antypas:2022asj}. 

Current searches for wave-like DM rely primarily on haloscope experiments~\cite{Sikivie:1983ip}, 
which use high-quality cavities to resonantly convert DM into detectable photons when the cavity frequency matches the DM mass. 
These efforts have achieved impressive sensitivity in the $1-10\;\mu$eV mass range (corresponding to GHz frequencies), with experiments like ADMX~\cite{PhysRevLett.127.261803,ADMX:2024xbv} and CAPP~\cite{CAPP:2024dtx} leading the field, supported by rapid technological developments~\cite{PhysRevLett.127.261803, HAYSTAC:2020kwv, PRXQuantum.3.030333, PhysRevLett.126.141302, Posen:2022tbs, DiVora:2022tro, Marconato:2023vqi}.

However, this experimental success covers only a tiny fraction of theoretically allowed parameter space,
leaving the eV mass range—corresponding to optical frequencies—almost completely unexplored~\cite{Chaudhuri:2014dla,Kahn:2016aff, Berlin:2022hfx,Giaccone:2022pke, Beurthey:2020yuq, Cervantes:2022epl, ALPHA:2022rxj,AlvarezMelcon:2020vee}. This represents both a critical gap in current search strategies and a substantial discovery opportunity. The eV range is especially compelling as it would complement searches for solar axions at future facilities like the International Axion Observatory~\cite{IAXO:2019mpb,IAXO:2020wwp}.

To address this unexplored regime, we propose a fundamentally different approach: integrated photonic resonator arrays fabricated on semiconductor wafers. 
To scale up our search we employ frequency multiplexing: rather than scanning one frequency at a time, we simultaneously monitor hundreds of potential DM masses using resonators with different frequencies coupled to a single readout system.
We exploit the fact that wave-like DM can resonantly excite EM modes in periodically structured optical resonators when their frequencies match the DM mass. These resonators couple to on-chip waveguides that guide signal photons to sensitive, low dark count single-photon detectors~\cite{SENSEI:2020dpa,Chou:2023hcc}, creating complete detection systems that can be mass-produced using semiconductor fabrication. 

Previous approaches to eV-scale DM detection have included optical reflector searches~\cite{BREAD:2021tpx} and dielectric stack experiments~\cite{Baryakhtar:2018doz,Chiles:2021gxk,Manenti:2021whp}.
However, our photonic approach offers unique advantages: more than hundreds of thousands of resonators can be fabricated and read out simultaneously on a single wafer. 
Unlike cavity experiments requiring individual assembly and tuning, photonic chips provide reproducible performance at scale, representing a new experimental paradigm that leverages the fabrication volume and maturity of photonics for fundamental physics discovery.

We demonstrate that this photonic approach can probe unexplored regions of axion and dark photon parameter space, including theoretically motivated areas such as the QCD axion band, while efficiently scanning across the $0.1-3$ eV mass range. Throughout this work we use natural units, i.e., $\hbar = c = \varepsilon_0 = \mu_0 = 1$, unless otherwise stated. %

\section{Dark Matter as an Electromagnetic Source}
\label{sec:electrodynamics_with_DM}
To understand how our photonic approach detects wave-like DM, we first establish how eV-scale dark matter behaves as a classical EM source, how this source can resonantly drive optical modes, and how periodic photonic structures enable efficient DM-photon conversion.

\subsubsection*{Wave-like Dark Matter Fields}
For DM masses in the eV range, the large occupation number per de Broglie volume—arising from the local DM density $\rho_{\rm D}\simeq 0.4$ GeV/cm$^3$~\cite{deSalas:2020hbh}—justifies treating the ultralight DM as a classical field.  
This wave-like DM oscillates almost monochromatically at a frequency determined by its mass $\mdm$, with small corrections from non-relativistic galactic motion. 

The DM field can be represented as a Gaussian random field\footnote{There is an equivalent representation of the DM field in terms of a discrete superposition of plane waves~\cite{Foster:2017hbq}.}~\cite{Krauss:1985ub} with spatial coherence extending over its de Broglie wavelength (see \cref{ap:incoherent-power}): 
\eqa{\label{eq:lambda}
    \ldb = \frac{2}{\mdm v} 
    \approx \, 0.4\;\mathrm{mm} \pfrac{\mathrm{eV}}{\mdm},
}
where $v \sim 10^{-3}c$~\cite{Evans:2018bqy} reflects typical DM velocities in our galaxy. This finite coherence scale will impose fundamental constraints on detector design, as we explore in \cref{sec:combining}.

The coherent nature of this field is what enables resonant detection strategies. Unlike conventional DM that would produce rare scattering events~\cite{Essig:2022dfa,Akerib:2022ort}, ultralight DM could create a persistent, weak EM source oscillating at optical frequencies, as we show below.

\subsubsection*{Axion and Dark Photon Electrodynamics}

When axions or dark photons couple to the SM photons and constitute the ambient DM, they modify Maxwell's equations by introducing an additional current density $\Jdm$ in Amp\`ere's law~\cite{Sikivie:1983ip,Graham:2014sha,Arias:2014ela} :
\beq\label{eq:maxwell}
 \nabla\times \vecbf{H} - \partial_t \vecbf{D} \approx \Jdm ,
\eeq 
where
\beq\label{eq:Jdm}
\Jdm=\begin{cases}
\gag \vecbf{B} \dot a(t,\vecbf{r})&\text{axion DM}, \\
\chi \mdm^2 \vecbf{A}'(t,\vecbf{r})& \text{dark photon DM}.
\end{cases}
\eeq
This current acts as a new EM source that can resonantly drive detectable fields in laboratory resonators tuned to the DM frequency. The form of this current depends on the specific DM model and determines the experimental requirements.

For \textbf{axion} DM, the interaction requires an external magnetic field $\vecbf{B}$ and produces a current $\Jdm= \gag \vecbf{B} \dot a(t,\vecbf{r})$,  
where $\gag$ is the axion-photon coupling and $\dot a$ represents the time derivative of the axion field.
For \textbf{dark photon} DM, the coupling is simpler, directly generating a current without external fields: 
$\Jdm=\chi \mdm^2 \vecbf{A}'(t,\vecbf{r})$, 
where $\chi$ is the kinetic mixing parameter and $ \vecbf{A}'$ is the dark photon field.

The oscillation frequency $\omega$ of the DM field depends on its velocity distribution, which is inferred from the Milky Way gravitational potential and from hydrodynamical simulations~\cite{Evans:2018bqy}. This distribution is well-modelled by a truncated Maxwellian with a disperion of $\sim 10^{-3}c$. As a result, the DM field modes are non-relativistic -- they oscillate nearly monochromatically at $\omega\simeq \mdm$ -- with frequency spread $\delta \omega/\omega \sim v^2/c^2 \sim10^{-6}$. This means that the DM  can resonantly drive EM modes in detectors whose frequency $\omega_R$ matches $\mdm$. 
For a critically-coupled resonator where internal losses equal readout losses, the extractable signal power is:
\beq
P_{\rm sig}=\frac{Q}{m_{\rm D}} \bar{J}_{\mathrm{D}}^2  |\eta|^2 V~,
\label{eq:generic_signal_power}
\eeq
where $Q$ is the quality factor of the mode on resonance, $V$ is the detector volume, $\eta$ quantifies the spatial overlap between the DM field and detector mode, and $ \bar{J}_{\mathrm{D}}^2$ captures the dependence on DM models:
\beq
\bar{J}_{\mathrm{D}}^2 =\frac{\rho_{\rm D}}{2 m_{\rm D}^2}\begin{cases}
\gag^2  m_{\rm D}^2 B^2 & \text{axion DM}\\
\chi^2 m_{\rm D}^4/3 & \text{dark photon DM}
\end{cases}.
\label{eq:dm_avg_current}
\eeq

\subsubsection*{Phase Matching and Resonant Enhancement}

The overlap factor $\eta$ in \cref{eq:generic_signal_power} is the key design parameter for optimizing DM-photon conversion efficiency. 
It quantifies how well the DM field couples to the EM mode, accounting for both the spatial profile of the resonator mode and the finite coherence of the DM field. 
For a resonant mode  $\vecbf{E}(\vecbf{x})e^{i\omega_Rt}$, we obtain (see \cref{ap:incoherent-power})
\beq\label{eq:eta}
|\eta|^2 \equiv
\frac{\int d^3 x\int d^3 x'E_{\parallel}^*(\vecbf{x})E_{\parallel}(\vecbf{x}') ~e^{-(\vecbf{x}-\vecbf{x}')^2/\ldb^2}}{ V\int d^3x~\varepsilon(\vecbf{x}) \left | \vecbf{E}(\vecbf{x})\right |^2}~,
\eeq
where $\varepsilon(\vecbf{x})$ is the spatially-varying dielectric function, $E_{\parallel}$ is $\vecbf{E}$ projected to the direction of $\vecbf{B}$ (for axions) or the dark photon polarization, and a Gaussian velocity distribution is assumed for DM. 

A fundamental challenge arises from momentum mismatch: dark matter is non-relativistic ($v \sim 10^{-3}c$) while optical photons are ultra-relativistic. Even within the coherence length, efficient coupling requires \textbf{phase matching} -- providing the additional momentum needed for energy-momentum conservation during the DM-photon conversion process.

Periodic modulation of the refractive index $\varepsilon(\vecbf{x})$, e.g. in a photonic crystal-like structure, such as grooved microrings~\cite{Wang:17} or patterned photonic crystal slabs~\cite{PhysRevLett.109.067401}, resolves this mismatch by supplying additional momentum. 
In particular, we find that DM-photon phase matching is possible using EM modes called Bound States in Continuum (BIC)~\cite{hsu2016bound}. These modes combine high quality factors and spatial profiles needed for DM coupling, achieving overlap factors of $0.01\sim 0.1$ (see \cref{ap:periodic_structures}) while maintaining excellent optical confinement. As a result, the DM field can efficiently excite BICs, making our photonic approach viable.

However, even with optimized phase matching, the signal power from individual resonators remains too small for practical detection of DM. A single resonator produces fewer than one signal photon per year for axion photon couplings near current experimental bounds (see \cref{sec:signal_rate}), motivating the wafer-based multi-resonator scaling strategies we develop in \cref{sec:combining}.

\section{Scaling to Discovery Sensitivity}
\label{sec:combining}

Having established how individual resonators couple to DM, we now address the critical challenge of combining signals from many resonators to surpass existing constraints from astrophysical observations. The fundamental obstacle lies in balancing signal coherence with detector scaling, which we resolve through a novel frequency multiplexing approach.

\subsubsection*{The Coherence Challenge}

The finite spatial coherence of DM creates a fundamental constraint that distinguishes our approach from traditional particle detection. For eV-scale DM, its coherence length is $\ldb\approx 0.4$ mm (cf. \cref{eq:lambda}), meaning that when detector dimensions exceed this scale, signals from different regions oscillate out of phase, leading to exponential signal suppression $\propto\exp[-({\rm detector~ size}/\ldb)^2]$ (cf. \cref{eq:eta}).

This coherence limitation means that naively combining many same-frequency resonators fails because their signals can interfere destructively when the total system size exceeds $\ldb$ (see \cref{ap:SI-combining}). Even within the coherence length, coupling multiple identical resonators to a single readout bus creates additional complications: signal photons can resonantly excite neighboring resonators before reaching the detector, leading to loss through internal dissipation.

\begin{figure*}[t]
     \centering
  \includegraphics[scale=0.28, trim =200 0 100 0, clip=true ]{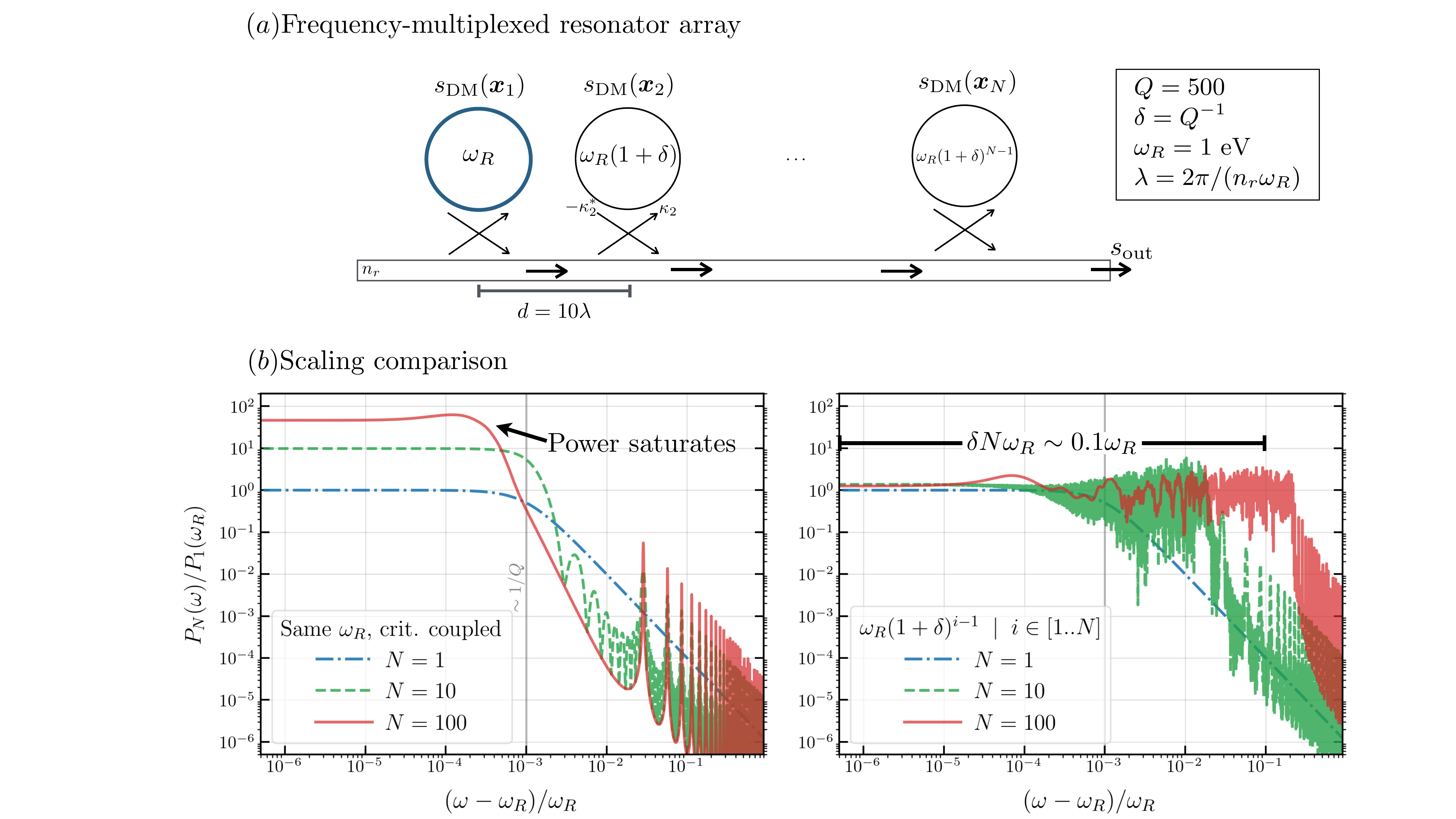}  
  \caption{$(a)$ Schematic of $N$ resonators with distinct frequencies $\omega_R$, $\omega_R(1+\delta)$, ..., $\omega_R(1+\delta)^{N-1}$ coupled to a single readout bus. DM sources $s_{\mathrm{DM}}$ driving each resonator have the same frequency, but potentially different phases, depending on the resonator separation $d$. The bus-resonator interaction is parametrized by coupling constants $\kappa$. Typical parameter choices are given in the box.
$(b)$ Signal power comparison (computed using coupled mode theory) demonstrates the scaling advantage of frequency multiplexing. \textbf{Left}: Same-frequency resonators ($\omega_R$, critically coupled) exhibit power saturation, preventing $N$-scaling, when the linear size of the array exceeds the DM coherence length. \textbf{Right}: Frequency-diverse resonators $\omega_R(1+\delta)^{i-1}$ eliminate inter-resonator interference, enabling independent operation and broadband coverage. 
This analysis provides the foundation for the experimental architecture presented in \cref{fig:setup}. }
\label{fig:series}
\end{figure*}

To quantify these effects, we analyze one-dimensional arrays of equally spaced resonators coupled to a common readout bus (see \cref{fig:series}) using coupled mode theory. 
Such arrays of coupled resonators are well-studied systems that enable photon dispersion engineering and enhanced non-linear effects, see, e.g., \cite{yariv_2023_z5c1t-ddk31, rabus2020integrated, heebner2008optical,KhurginTucker2009} and references therein. Compared to standard applications of these arrays, our source $\Jdm$ is not necessarily coherent over the entire system. Additionally, in standard applications the resonator array is usually excited from a single input port (one side of the bus) and the output is measured at the other end. For DM detection each ring acts as a possibly incoherent source of waves in the bus propagating in both directions, which requires a novel reformulation of the coupled mode theory~\cite{784592}. We formulate this theory and summarize our main findings below, with detailed analytic and numerical solutions provided in \cref{sec:cmt}.

\subsubsection*{Coupled Mode Theory}
Consider the setup sketched in the upper panel of \cref{fig:series} with $N$ resonators of (potentially) different resonant frequencies $\omega_{R_l}$ coupled to a single bus. 
DM could source a standing wave in resonators with $\omega_{R_l}$ close to the DM mass.
The DM sourced photon may leak to the environment due to a finite intrinsic quality factor of the resonator $Q_{\rm i}$ or to the receiver bus due to a designed bus-resonator coupling $\kappa$. When leaking to the bus, the signal photon can travel in either direction and can potentially excite neighboring resonators. 
Therefore, the $l$th resonator amplitude $a_l$ receives contributions from both DM and left or right-moving bus photons.

Such a coupled system can be described by Coupled Mode Equations, which in frequency space read (following the notation of Ref.~\cite{784592})
\beq
\left[i(\omega_{R_l} - \omega) - \tau^{-1}_l\right] a_l + \kappa_{1,l} s_{+1,l} + \kappa_{2,l} s_{+2,l} = s_{\mathrm{DM},l},
\label{eq:cmt_equation}
\eeq
where $s_{+1,l}$ ($s_{+2,l}$) is the photon amplitude in the bus arriving from the left (right) at the $l$th resonator, and $\kappa_{i,l}$ is the corresponding 
bus-resonator coupling.
The amplitudes $a_l$ ($s_{+i,l}$) are normalized such that their time averages $\langle a_l(t)^2\rangle $ ($\langle s_{+i,l}(t)^2\rangle$) give the energy stored in the $l$'th resonator (power flowing through the bus at the $l$'th resonator).
$s_{\mathrm{DM},l} \equiv s_{\mathrm{DM}}(\omega,\vecbf{x}_l)$ represents the DM source at the position of the $l$th resonator:
\beq
s_{\mathrm{DM},l} =  -\frac{\omega}{2\omega_{R_l}} \frac{\int d^3 r \vecbf{E}_l \cdot\Jdm^+(\omega,\vecbf{x}_l)}{\sqrt{\int d^3 r \varepsilon \vecbf{E}_l^2}},
\label{eq:dm_source_in_coupled_mode_equations}
\eeq
where $\vecbf{E}_l$ is the spatial profile of the $\omega_{R_l}$ mode in the $l$th resonator, and $\varepsilon$ is the dielectric constant of the resonator. \cref{eq:dm_source_in_coupled_mode_equations} can be obtained by Fourier transforming~\cref{eq:maxwell,eq:Jdm} and taking the real components of EM fields near $\omega \approx \omega_{R_l} > 0$;  $\Jdm ^+$ is the positive frequency part of the DM source. We have assumed that the DM field is spatially-coherent over a single resonator, while allowing for different phases at different resonators; as a result $\Jdm^+$ only 
depends on the position of the resonator, $\vecbf{x}_l$, but not on the integration over the resonator volume.

The bus power amplitudes $s_{+i,l}$ couple all of the resonators together; in the weak coupling limit energy conservation allows us to express these quantities in terms of the 
mode amplitudes of other resonators on the bus:
\begin{equation}
  s_{+1, l} = e^{-i \omega n_r d}\begin{cases}
  s_{+1, l-1} - \kappa_{2,l-1}^* a_{l-1} & l > 1\\
  0 & \text{otherwise} 
  \end{cases}
  \label{eq:right_propagating_power_amp}
\end{equation}
and 
\begin{equation}
  s_{+2,l} = e^{+i\omega n_r d}\begin{cases}
  s_{+2,l+1} - \kappa_{1,l+1}^* a_{l+1} & N> l \geq 0\\
  0 & \text{otherwise} 
  \end{cases}
  \label{eq:left_propagating_power_amp}
\end{equation}
where $n_r$ is the effective refractive index of the receiver bus, $d$ is the spatial separation between the neighboring resonators. The exponential factors represent the phase accumulated as the signal propagates through the bus between resonators. 

$\tau_l$ is the lifetime of the photon in the $l$th resonator and is determined by the intrinsic $Q_{\rm i}$ of each resonator and the bus-resonator couplings. The symmetry of this configuration implies that the forward and backward bus-resonator couplings are equal: 
$\kappa_l\equiv\kappa_{1,l}=\kappa_{2,l}$. 
Thus, the total decay rate $\tau^{-1}_l$ can be written as
\beq
\tau^{-1}_l = \tau_{\mathrm{i},l}^{-1}+\tau_{\mathrm{e},l}^{-1} 
\eeq
where $\tau_{\mathrm{e},l}^{-1}= |\kappa_l|^2$ is the partial width of the photon decaying into bus modes propagating in either direction and
$
\tau_{\mathrm{i},l}^{-1} = \frac{\omega_{R_l}}{2 Q_\mathrm{i}}
$ is the intrinsic loss rate. Similar to $Q_{\rm i}$, one introduce external quality factor $Q_\mathrm{e}$ such that  $|\kappa_l|^2 \equiv \frac{\omega_{R_l}}{2Q_\mathrm{e}}$.
Note that the decay rate of the mode energy $a_l(t)^2$ is $2\tau_l^{-1}$. Lastly, the total or loaded 
quality factor includes both intrinsic and extrinsic losses is given by 
\beq
Q^{-1} =  Q_\mathrm{i}^{-1} + Q_\mathrm{e}^{-1}
\eeq
and determines the physical width of the bus-coupled resonator. 

Note that while the sketch in \cref{fig:series}a shows the resonators as rings, coupled mode theory accommodates any resonator form factor for an appropriate choice of $\vecbf{E}_l$ and coupling constant $\kappa$.

We want to know the steady-state power that the system emits through the bus to the left or right of all the resonators; iterating ~\cref{eq:right_propagating_power_amp,eq:left_propagating_power_amp} 
this output power amplitude is given by
\beq
s_{\rm out}(\omega) =\sum_{l=1}^N-\kappa^*_l a_l e^{-i\omega n_rd(l-1)} \equiv -\vecbf{t}\cdot \vecbf{a} 
\label{eq:output_power_amplitude_one_direction}
\eeq
where $
\vecbf{a}  =(a_1,a_2,\cdots,a_N)$, and $
\vecbf{t} =\left(\kappa_1^*,\kappa_2^* e^{-i\omega n_r d},\cdots,\kappa_N^*e^{-i(N-1)\omega n_r d}\right)$. 
By inverting the linear system in~\cref{eq:cmt_equation}, we obtain
\begin{widetext}
\begin{equation}
\label{eq:M}
\vecbf{a} = M^{-1}\vecbf{s}_{\mathrm{DM}}= \begin{bmatrix}
    i(\omega-\omega_{R_1}) + \tau_1^{-1} & \kappa_1\kappa^*_2e^{-i\omega n_rd}   & \cdots & \kappa_1\kappa^*_N e^{-i(N-1)\omega n_rd} \\
    \kappa_1\kappa^*_2 e^{-i\omega n_rd} & i(\omega-\omega_{R_2})  + \tau_2^{-1} & \cdots & \kappa_2\kappa^*_N e^{-i(N-2)\omega n_rd}  \\
    \vdots & \vdots & \ddots & \vdots \\
    \kappa_1\kappa_N^*e^{-i(N-1)\omega n_rd}  & \kappa_2\kappa_N^*e^{-i(N-2)\omega n_rd}  & \cdots & i(\omega-\omega_{R_N}) + \tau_N^{-1}
\end{bmatrix}^{-1} \vecbf{s}_{\mathrm{DM}}
\end{equation}
\end{widetext}
where $\vecbf{s}_{\mathrm{DM}} = (s_{\mathrm{DM},1},\dots,s_{\mathrm{DM},N})$. 
The power emitted at the end of the bus, averaged over time $T$, is therefore
\beq
P_{\mathrm{sig}} = \frac{2}{T}\int_0^\infty \frac{d\omega}{2\pi} |s_{\rm out}(\omega)|^2,
\label{eq:sig_power_in_one_direction}
\eeq
where 
\beq
\begin{split}
|s_{\rm out}(\omega)|^2
=\sum_{ijkl} t_i  M^{-1}_{ij} s_{\mathrm{DM},j}s_{\mathrm{DM},k}^*  {{M}^{-1}}^{\dagger}_{kl} {t}^*_l~.
\end{split}
\label{eq:fourier_power_amp_squared}
\eeq
Note that this spectral distribution is for signal radiated into one direction of the bus. If both ends of the waveguide are read out, then the total signal power is doubled.

 We see that the output power is given by the magnitude-squared of an overlap between a readout mode $\vecbf{t}$ and a DM source vector $\vecbf{s}_{\mathrm{DM}}$. A similar result is obtained in \cref{ap:SI-combining}  for the case of $N$ non-interacting resonators. 
However, \Eq{eq:fourier_power_amp_squared} now includes the effects of interactions via the off-diagonal entries in $M$, proportional to $\kappa_i \kappa_j^*$.  The $\kappa_l$ depend on the geometry, material properties and mode profiles of the resonator and bus.

Since the resonator spacing can be larger than the DM coherence length, the DM velocity dispersion is no longer negligible and $|s_{\rm out}(\omega)|^2$ will depend on 
the relative phases of the DM field at the different resonators. 
To obtain the expected steady-state output power we can average $|s_{\rm out}(\omega)|^2$ over possible 
DM field realizations for a given DM velocity distribution $f(\vecbf{v})$. 
Treating DM as a classical random Gaussian field, we find from \cref{eq:dm_source_in_coupled_mode_equations}
\beq
\begin{split}
\langle s_{\mathrm{DM},j}& s_{\mathrm{DM},k}^* \rangle  = \frac{\omega^2}{4\omega_{R_j}\omega_{R_k}} \sqrt{V_j}\eta_{1,j} \sqrt{V_i}\eta_{1,k}^* \bar{J}_{\mathrm{DM}}^2 \\
 &\times \int d^3 v f(\vecbf{v}) e^{ i m\vecbf{v}\cdot (\vecbf{x}_j-\vecbf{y}_k) }  2\pi T \delta (\omega - m) 
\end{split}
\label{eq:dm_source_ensemble_average}
\eeq
where $\eta_{1,j}= \frac{V_j^{-1}\int d^3 r \vecbf{E}_j \cdot \hat{\vecbf{n}}}{\sqrt{V^{-1}_j\int d^3 r \varepsilon \vecbf{E}_j^2}}$ are the single resonator overlap factors 
and $\bar{J}_{\mathrm{DM}}^2$ is given in \cref{eq:dm_avg_current}.
The velocity distribution above is normalized such that $\int d^3 v f(\vecbf{v})=1$; as in \cref{sec:electrodynamics_with_DM} we will take a Gaussian $f(\vecbf{v})$ for simplicity:
\beq\label{eq:gaussian}
f({\vecbf{v}})=\frac 1{\pi^{3/2}v_0^3}e^{-(\vecbf{v}+\vecbf{v}_{\odot})^2/v_0^2},
\eeq
which allows to us to evaluate the velocity integral in~\cref{eq:dm_source_ensemble_average} explicitly:
\beq
\begin{split}
\int d^3 v f(\vecbf{v}) \exp i\left[m\vecbf{v}\cdot (\vecbf{x}_j-\vecbf{x}_k)\right]  = \\
 \exp\left( -\frac{1}{4}m^2 v_0^2  (\vecbf{x}_j-\vecbf{x}_k)^2  -i\left[m\vecbf{v}_\odot \cdot (\vecbf{x}_j-\vecbf{x}_k)\right]\right)
\end{split}~.
\label{eq:dm_velocity_integral_gaussian}
\eeq
For the linear arrangement of resonators depicted in \cref{fig:series}, $(\vecbf{x}_j-\vecbf{x}_k)^2=(d|j-k|)^2$.
Since $v_0\sim |{\bf v}_{\odot}|\sim 10^{-3}c$, $\langle s_{\mathrm{DM},j} s_{\mathrm{DM},k}^* \rangle$ becomes exponentially suppressed when $\omega d |j-k|\gg 10^3$. 
In this limit, the DM source correlation matrix, \cref{eq:dm_source_ensemble_average}, is proportional to the identity, corresponding to each resonator being incoherent with respect to its neighbours. 
However, as we will see in \cref{sec:projected_sensitivity}, we will need to maximize the number of resonators on a single bus while demanding that each resonator is of order of the coherence length of DM (to maximize its volume). Thus we will be concerned with $d\sim 1/(m v_0)$ and some level of source correlation is inevitable between neighbouring resonators.

In \cref{sec:cmt} we use this framework to find analytic and numerical solutions for various choices of number of coupled resonators, quality factors and resonator separations. We give a summary of our main results below. Importantly, we find that for $N$ resonators with identical frequencies $\omega_R$ and quality factors and nearest neighbor separation $d$, the system dynamics reveal a critical distinction between coherent and incoherent DM sources:
 \begin{itemize}
 \item \emph{Coherent case}. Here all all resonators lie within one coherence length, requiring $v \omega_R d N \ll 1$. We find that for $N$ resonators on resonance: 
\beq
P_{\mathrm{sig},\;{\rm coherent}}=\frac{Q_{\rm i}}{\mdm}\frac{N^2\beta}{(1+N\beta)^2}\bar{J}_{\mathrm{DM}}^2  |\eta_1|^2 V~, \label{eq:sigP_N}
\eeq
where $\beta = \tau^{-1}_{\mathrm{e}}/\tau_\mathrm{i}^{-1}$. 
The optimal coupling $\beta=1/N$ yields $N$-fold signal enhancement compared to the single resonator result (\cref{eq:generic_signal_power} applied to one resonator), corresponding to ``critical coupling'' for the entire array.
\item \emph{Incoherent case} is relevant for realistic array sizes $Nd > \ldb$, which is needed to surpass existing constraints on DM parameter space. Our analysis reveals that the signal power at resonance does not scale with $N$. This occurs because the overlap factor $\eta$ suffers exponential suppression when the total array extent exceeds the DM coherence length, as described below \cref{eq:dm_velocity_integral_gaussian}. 
This represents a fundamental physics limitation, not an engineering challenge. The importance of this effect is illustrated in the left panel of \cref{fig:series}b. As the number of coupled resonators increases, the total signal power does not scale linearly with $N$. Moreover, the bandwidth of the detector decreases.
 \end{itemize}

 We conclude that for realistic array sizes, it is not feasible to achieve $N$-fold power enhancement at a single frequency.

\subsubsection*{Frequency Multiplexing Solution}

The key insight is that resonators at different frequencies can be addressed individually, akin to add-drop multiplexing strategies in telecommunications \cite{little:1997JLT}. Rather than attempting to achieve $N$-fold power enhancement at a single frequency -- which fails due to coherence constraints -- we monitor $N$ different frequencies simultaneously, each with single-resonator sensitivity.
This \textbf{frequency multiplexing} strategy transforms the scaling trade-off: instead of seeking enhanced signal power at one frequency, we achieve enhanced bandwidth coverage while maintaining sensitivity. The approach is illustrated in the right panel of \cref{fig:series}b.

Consider $N$ resonators with distinct frequencies $\omega_R, \omega_R(1+\delta), \cdots, \omega_R(1+\delta)^{N-1}$ coupled to a single readout bus, where $\delta\sim Q^{-1}$. Each resonator targets a different DM mass, with frequency spacing chosen to provide gapless coverage across the monitored bandwidth. The DM sources $s_{\rm DM}(\vecbf{x}_i)$ driving each resonator (related to the spatial integral of $\Jdm$) are spatially incoherent when separated by more than $\ldb$. However, since no two resonators share the same frequency, each resonator operates independently with signal power given by \cref{eq:generic_signal_power}. While each frequency maintains only single-resonator sensitivity, the total monitored bandwidth grows linearly with $N$. For $N \ll Q$, the bandwidth 
$
\Delta\omega\approx \delta N\omega_R = N\omega_R/Q 
$ which is borne out in the numerical solution of the signal power in \cref{fig:series}b.

This frequency multiplexing paradigm represents a fundamental shift from traditional cavity-based DM searches. Rather than precisely tuning a single bulky high-$Q$ cavity to match an unknown DM mass—requiring sequential scanning and re-tuning (see, e.g.,~\cite{Adair:2022rtw,PhysRevLett.127.261803}), we create a broadband `spectrum analyzer' that simultaneously monitors hundreds of potential DM masses.
This approach is  
naturally suited to the unknown  nature of DM masses. Moreover, it leverages the key advantage of integrated photonics: the ability to fabricate over hundreds of thousands of precisely-controlled resonators on a single wafer using semiconductor fabrication techniques.

\section{Experimental Architecture and Projected Sensitivities}
\label{sec:projected_sensitivity}

Building on the frequency multiplexing architecture described in \cref{sec:combining}, we now present a conceptual experimental design and project its sensitivity to ALP and DP DM candidates. 
Our approach leverages integrated photonics to enable substantial sensitivity improvement across the 0.1-3 eV mass range.

\subsubsection*{Experimental Architecture}

\begin{figure}
     \centering
 \includegraphics[clip =true, trim={300 100 450 50}, width=8.5cm]{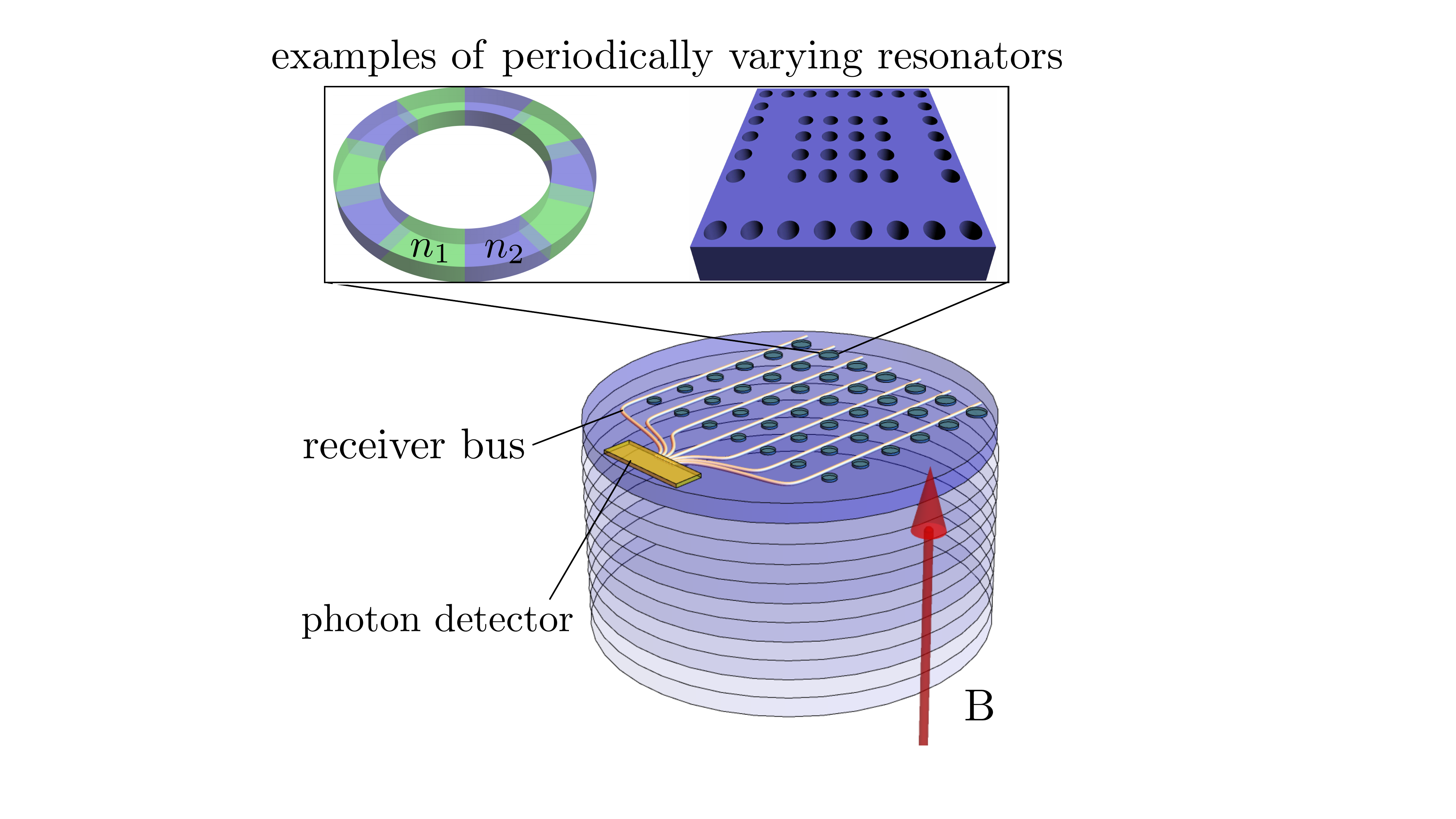}
  \caption{Conceptual architecture for wafer-scale DM detection. Periodically-modulated resonators (inset, with discretely varied refractive indicies $n_1$ and $n_2$) provide phase matching for efficient DM-photon coupling. Frequency-multiplexed readout enables parallel monitoring of hundreds of potential DM masses, while low dark count single-photon detectors provide the noise performance needed for discovery. 
 }\label{fig:setup}
\end{figure}

\cref{fig:setup} illustrates the conceptual design of the proposed detector. 
Each photonic chip contains arrays of resonators with distinct frequencies $\omega_R, \omega_R(1+\delta), ..., \omega_R(1+\delta)^{N-1}$, where $\delta = Q^{-1}$ ensures complete spectral coverage without gaps. 
The system integrates three critical features: (1) phase matching through periodic structures, (2) frequency-multiplexed readout implementing the strategy from Section III, and (3) scalable photonic fabrication. As shown in \cref{sec:signal_rate}, a single 15 cm (6 inch) diameter wafer can accommodate $\sim10^6$  resonators for eV-scale frequencies, with each chip containing arrays targeting different frequency bands. Multiple wafers can be combined to instrument larger volumes, for instance with fiber-optic interconnects or hybrid integration, with the total active volume limited only by available magnet bore size for axion DM searches or simply laboratory space for DP DM searches.

Critically, this approach transforms DM detection from precision tuning of individual cavities to parallel processing of mass-produced components. The same fabrication techniques used for telecommunications photonics, with established yield, precision, and cost metrics, directly enable fundamental physics exploration. This is a qualitative shift from traditional approaches that require custom assembly and individual optimization of each detector element.

\begin{table}[t]
    \centering
\begin{tabular}{ccccc}
    \hline
  Detector & Range ($\eV$) & DC ($\sec^{-1}$) & Size ($\micron^2$) & Refs.  \\
    \hline
  SNSPD & $0.1  - 1.12 $ & $ 6 \cdot 10^{-6} $ & $ 400 \times 400  $ & \cite{Hochberg:2021yud,Chiles:2021gxk}\\
  CCD & $ 1.12 - 3  $ & $ 10^{-9}  $ & $ 15\times15  $  & \cite{SENSEI:2023zdf}\\
    \hline
\end{tabular}
\caption{
Detector technologies considered here.
Columns show the frequency range , dark count rate and physical collecting area per element, and relevant references.  The threshold energy $1.12\; \eV$ is the band gap of silicon. Signal photons below (above) the band gap are detected with SNSPDs (CCDs).
}
\label{tab:dc}
\end{table} 

Furthermore, the architecture naturally interfaces with two complementary single-photon detector technologies optimized for different energy ranges. As shown in \cref{tab:dc}, superconducting nanowire single-photon detectors (SNSPD) provide optimal performance for the 0.1-1.12 eV range, while skipper charge-coupled devices (CCD) achieve superior performance above 1.12 eV. 
\\

\subsubsection*{Sensitivity Projections}\label{sec:reach}

The achievable sensitivity depends on three key factors: total active volume, detector background rates, and integration time. For discovery-level sensitivity, we require a signal-to-noise ratio SNR $\geq 1$ with
\eqa{
  \tx{SNR} = \frac{\Gamma_\tx{sig} t_\tx{int}}{\Max \lb 1, \Gamma_\tx{bkg} t_\tx{int} \rb^{1/2} }~.\label{eq:snr}
}
$\Gamma_\tx{sig}$ $ (\Gamma_\tx{bkg})$ is the signal (background) rate, $t_{\rm int}$ is the integration time.
The Max function captures two distinct experimental regimes with different scaling behavior. 

In the \textbf{background-free} regime (when $\Gamma_\tx{bkg} t_{\rm int}\ll1$), detector backgrounds are negligible the SNR scales linearly with integration time. 
In the \textbf{background-limited} regime (when $\Gamma_\tx{bkg} t_{\rm int}\gg 1$), detector backgrounds dominate and the signal must compete against statistical fluctuations in the background rate. The SNR scales as $\sqrt{ t_{\rm int}}$. This regime shows the characteristic square-root degradation of background-limited searches. 

\begin{figure*}[t]
\centering
\includegraphics[width=12cm]{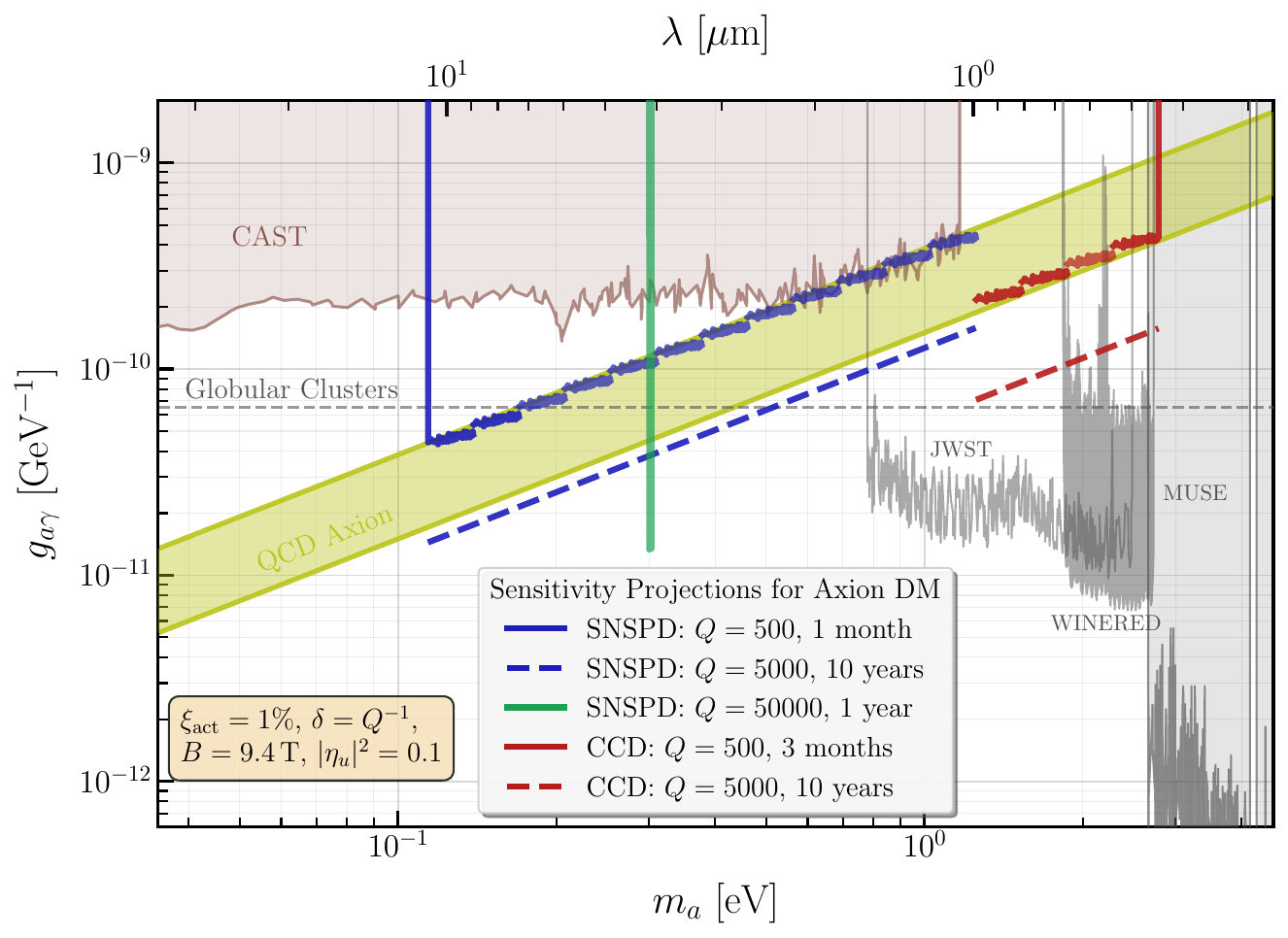}
\caption{Projected axion-photon coupling sensitivity assuming SNR=1 in~\cref{eq:snr}. The bold colored lines show future sensitivity projections for different experimental configurations: SNSPD-based detectors (blue/green) covering 0.1--1.12~eV and CCD-based detectors (red) covering 1.12--2~eV mass ranges. Our approach can probe couplings $g_{a\gamma} < 10^{-10}$~GeV$^{-1}$, reaching into the theoretically motivated QCD axion band (yellow). Solid lines represent near-term projections with $Q=500$ resonators, while dashed lines show potential improvements with higher-$Q$ resonators over longer integration times. The green line demonstrates the exceptional sensitivity achievable in a narrow mass range with $Q=50000$ resonators and 1-year integration time. For context, current limits from CAST~\cite{CAST:2017uph}, stellar evolution constraints~\cite{Ayala:2014pea, Dolan:2022kul} (gray dashed line), and telescope-based searches~\cite{Todarello:2023hdk, Grin:2006aw, Janish:2023kvi, Yin:2024lla} (MUSE, JWST, WINERED) are shown as well, taken from the repository AxionLimits~\cite{AxionLimits}.
}\label{fig:axion_projection}
\end{figure*}

\begin{figure}
\centering
\includegraphics[width=8cm]{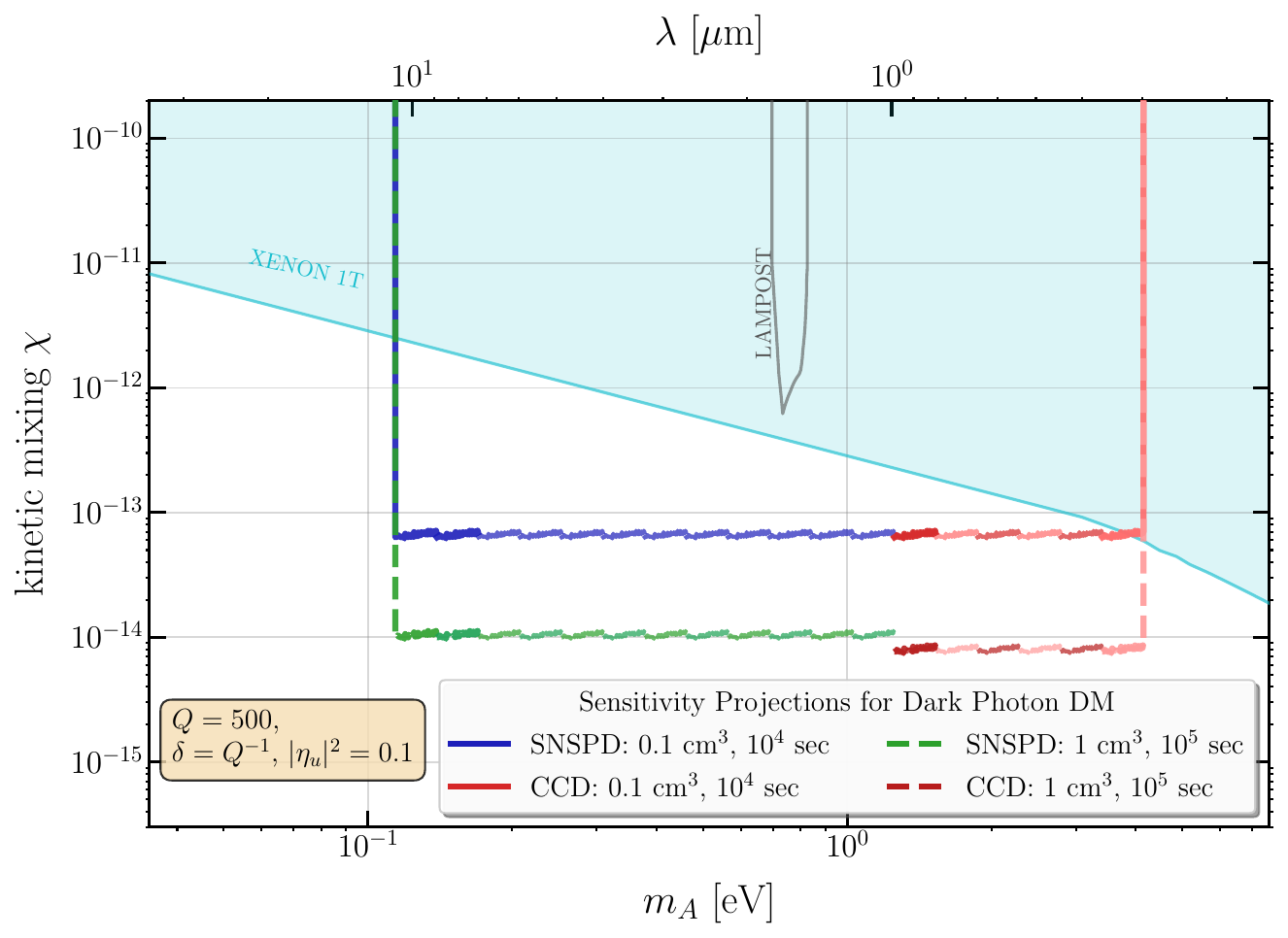}
\caption{
Projected kinetic mixing sensitivity for DP DM detection using SNSPD-based detectors (blue/green) and CCD-based detectors (red) with two different experimental scales. 
Also shown are the existing limits from solar dark photon searches (Xenon1T~\cite{An:2020bxd}) and direct detection experiments (LAMPOST~\cite{Chiles:2021gxk}). 
}\label{fig:dark_photon_projection}
\end{figure}

The signal rate scales as $\Gamma_{\text{sig}} \propto N_{\text{bus per wafer}} \times N_{\text{wafers}} \times \Gamma_1$, with $\Gamma_1$ the single-resonator rate. The background rates $\Gamma_{\text{bkg}}$ depend on the detector technology and total detector area. In \cref{sec:bkg} we estimate the typical dark count rate assuming an active volume of $1000$ cm$^3$ made up of 15 cm diameter wafers. 
\cref{eq:snr} provides the sensitivity framework used to generate the projections in \cref{fig:axion_projection} and \cref{fig:dark_photon_projection}, covering the 0.1-3 eV mass range for ALP and DP DM.

\paragraph*{Axion and ALP DM}
For axion DM searches requiring external magnetic fields, the sensitivity is limited by available magnet bore volume. Assuming complete instrumentation of a 9.4 T MRI magnet bore with our photonic arrays, for $\mdm=0.25$ eV, the axion-photon coupling sensitivity is: 
\begin{widetext}
\begin{equation}
\label{eq:resonant_axion_sensitivity_estimate}
  g_{a\gamma}^* \approx  \frac{2\mdm}{\sqrt{\rho_{\rm D}}}\sqrt{\frac{N}{B^2 V_{\rm act} Q|\eta_u|^2}} \left(\frac1{t_{\rm int}\Gamma_{\rm bkg}^{-1}}\right)^{1/4}
 \sim \frac{10^{-10}}{\GeV}\left(\frac{1~\rm month}{t_{\rm int}}\right)^{1/4}
\left(\frac{13.3~\rm sec}{\Gamma_{\rm bkg}^{-1}}\right)^{1/4}\left(\frac{67~\rm kJ}{ B^2V_{\rm act}}\right)^{1/2}\left(\frac{N}{100}\right)^{1/2}~,
\end{equation}
\end{widetext}
where an active volume fraction $\xi_{\text{act}} \sim 1\%$ of magnet bore volume containing millions of periodic resonators is assumed.  
The scaling of the sensitivity follows from our assumption that same-frequency resonators do not interfere when placed on separate buses; as a result the total signal rate at a particular frequency is simply proportional to $V_{\rm act}/N$. 
\cref{fig:axion_projection} shows the projected sensitivities for axion-photon coupling $\gag$ across the 0.1-2 eV mass range. 
Since different cosmological and particle physics models can predict other mass-coupling relations (see, e.g., ~\cite{Blinov:2019rhb} and references therein), it is important to probe as much of the $\mdm - \gag$ plane as feasible. 
To survey the 0.1-2 eV mass range, a scanning strategy is required and is ultimately determined by the detector technology. 
The silicon bandgap at 1.12 eV creates a natural division:
\begin{itemize}
\item
Phase 1 (0.1-1.12 eV): 12 sequential runs using SNSPDs (blue solid), each requiring 1 month  given $\Gamma_{\rm bkg}^{-1}\approx 13.3$ sec.
\item
Phase 2 (1.12-3 eV):  4 sequential runs using CCDs (red solid), each requiring 3 months with $\Gamma_{\rm bkg}^{-1}\approx  112.5$ sec .
\end{itemize} 
The complete coverage of the 0.1-3 eV range requires approximately 2 years.
The results demonstrate that this approach can probe couplings well into the theoretically motivated QCD axion band and is competitive compared to existing stellar evolution bounds (Globular Clusters) and helioscope searches (CAST). 
Higher quality factors $Q = 5000$ (shown as dashed lines) could improve sensitivity further, though at the cost of longer survey times due to narrower individual resonance bandwidths.
 
\paragraph*{Dark Photon DM}
While ALP DM searches demonstrate the ultimate discovery potential of our approach, DP DM detection offers a more accessible entry point with simpler experimental requirements. 
Unlike searches for ALPs, dark photon experiments require no external magnetic field (see \cref{eq:dm_avg_current}). 
Therefore, the key experimental parameter is simply the active volume of the detector.
Unlike the background-dominated regime in \cref{eq:resonant_axion_sensitivity_estimate}, DP DM search can be sensitive already in a \textbf{background free} regime, where the coupling sensitivity is approximately 
\beq
\begin{split}
&\chi^*  \approx \sqrt{\frac{12 N }{Q\rho_{\rm D} V_{\rm act} t_{\rm int} |\eta_u|^2}} \\
&\approx 10^{-13}\left(\frac{10^4\;\rm sec }{t_{\rm int}}\right)^{1/2}
\left(\frac{0.1\;\mathrm{cm}^3}{V_{\rm act}}\right)^{1/2}
\left(\frac{N}{100}\right)^{1/2} .
\end{split}
\label{eq:resonant_dark_photon_sensitivity_estimate}
\eeq 
\cref{fig:dark_photon_projection} presents our projected sensitivity to the kinetic mixing parameter $\chi$ for DP DM. The solid and dashed lines show projections for different active volumes (0.1 cm$^3$ and 1 cm$^3$) and integration times (10$^4$ sec and 10$^5$ sec). 
The comparison with current bounds from LAMPOST and Xenon1T shows that photonic detection opens substantial new parameter space for DPDM searches.
Importantly, \cref{fig:dark_photon_projection} demonstrates that even modest-scale proof-of-principle experiments with 0.1 cm$^3$ active volume could achieve competitive sensitivity, making dark photon searches a natural first step for validating this detection concept before scaling to larger axion DM search experiments requiring magnetic field infrastructure.

\subsubsection*{Technical Implementation}

Target performance requires only modest fabrication demands and is consistent with current low-loss integrated photonics standards across multiple platforms including silicon, silicon nitride, or lithium niobate, with the latter approach allowing electro-optically tuned resonators.
Fabrication precision is based on deep-UV or electron beam lithography \cite{madou2018fundamentals}, and state of the art Q-factors generally exceed one million \cite{han2024design, puckett2021422,zhu2024twenty}.
Our approach requires only relatively modest $Q = 500-5000$ for optimized DM readout. 
This can be achieved in index-modulated, e.g., periodically etched or cladded microrings \cite{Wang:17} or compact resonator designs, such as photonic crystals \cite{meade:2008textbook}, as discussed. 
Such designs both enhance DM photon coupling while ensuring a small footprint for scaling-up. Higher quality factor resonators can be used to achieve sensitivity to smaller couplings, at the price of very narrow mass coverage. This can be relevant if a broadband search detects a signal at a specific frequency. This is illustrated in \cref{fig:axion_projection} by the $Q=5\times 10^4$ reach estimate.

Frequency precision demands thermal stability of $\pm0.1$ K which is achieved either through conventional packaging or integrated heaters.
Furthermore, ultra-low loss readout ($<$1 dB) to detectors can be achived either by incorporating detectors on chip or using mode converter strategies, such as those employed by leading quantum computing companies \cite{psiquantum2025manufacturable}.
Exciting future implementations could incorporate active resonator tuning for real-time frequency scans, phase-sensitive or phase- and temporally modulated readout~\cite{Knirck:2018knd,Foster:2020fln} for DM velocity measurements, and quantum-enhanced detection protocols~\cite{PRXQuantum.2.040350,Chen:2023ryb,PRXQuantum.3.030333}. 
Such a temporally dynamic multi-resonator readout approach cannot be easily realized using conventional bulky methods. 
The relatively modest Q-factors also suggest that our proposals can readily benefit from emerging photonic platforms as they come available, for instance wide bandgap materials, such as fluorinated crystals \cite{zhang2024229thf4}, enable testing different DM mass assumptions thus gaining access to larger parameter space.

\section{Conclusion}

We have demonstrated that integrated photonics offer a transformative approach to eV-scale DM detection. Our frequency multiplexing strategy relaxes the tension between signal coherence and detector scaling, enabling simultaneous monitoring of broad ranges of potential DM masses rather than sequential frequency scanning.
The projected sensitivities of axion-photon couplings down to $10^{-11} {\rm GeV}^{-1}$ and dark photon-photon mixing below $10^{-14}$—reach well into theoretically motivated parameter space including the QCD axion band, across the entire 0.1-3 eV mass range.


Our analysis provides a clear experimental roadmap: DP DM searches offer an ideal starting point requiring no external infrastructure, while success would justify scaling to discovery-level axion experiments. The manufacturing scalability and maturity of integrated photonics--hardware that is now suporting the proliferation of data centers for internet and artificial intelligence-- makes this approach immediately implementable, potentially opening a new window on dark matter within the next decade.

\begin{acknowledgments}
We thank Yoni Kahn, Alex Millar, Masha Baryakhtar, Junwu Huang, Daniel Egana-Ugrinovic, Asher Berlin, Nick Rodd and Albert Stebbins for helpful discussions. 
This work is supported by the DOE QuantISED program through the theory  consortium "Intersections of QIS and Theoretical Particle Physics" at Fermilab.
 Fermilab is operated by the Fermi Research Alliance, LLC under Contract DE-AC02-07CH11359 with the U.S. Department of Energy. RH and RJ are also supported by the U.S. Department of Energy, Office of Science, National Quantum Information Science Research Centers, Superconducting Quantum Materials and Systems Center (SQMS) under contract number DE-AC02-07CH11359. CG and RH also acknowledge the Aspen Center for Physics for its hospitality where part of this work is done, which is supported by National Science Foundation grant PHY-1607611. NB is supported by Natural Sciences and Engineering Research Council of Canada (NSERC).
\end{acknowledgments}

\appendix

\section{Dark Matter Field and Signal Coherence}
\label{ap:incoherent-power}
In this section we relate the (known) DM velocity distribution to the statistical properties of the DM source 
that appears in \cref{sec:electrodynamics_with_DM}. We use this to derive the generic expression for DM signal power including 
coherence effects that appears in \cref{eq:generic_signal_power,eq:eta}. Here we refer to the DM as axions, but completely analogous reasoning applies to dark photon DM. 
\subsection{DM as a Random Field}
It is known in optics that the density operator
\beq\label{eq:densityop}
\hat\rho=\Pi_i \hat\rho_i=\Pi_i \sum_{n_i} \frac {\langle n_i\rangle^{n_i}}{\left(\langle n_i\rangle+1\right)^{n_i+1} }|n_i\rangle \langle n_i|
\eeq
where the summation runs over all the modes, applies not only to the thermal photon distribution but also a wide range of excitations in which the statistical properties of the light are suitably random \cite{Loudon}. Since the axion dark matter is stochastic, \cref{eq:densityop} could be a good description. Given an axion field operator
\beq\label{eq:phi}
\hat\phi({\bf x},t)= \sum_{l=1}^N \frac1{\sqrt{2\omega_l V}}\left( \hat a_ l e^{i{\bf x}\cdot {\bf k}_l-i\omega_l t}+h.c.\right)=\hat \phi^++\hat \phi^-
\eeq
where $\omega_l\simeq m+\frac12 mv_l^2$, we can compute observables using $\langle \mathcal{O}(\phi)\rangle={\rm Tr}\left(\rho\mathcal{O}(\phi)\right)$. If $n, \langle n\rangle\to\infty$, the probability $P(n)$ approximately takes the following form 
\beq
P(n)=\frac 1{\langle n\rangle +1}\left(1- \langle n\rangle^{-1}\right)^{n}\approx \frac 1{\langle n\rangle}e^{-n/\langle n\rangle}.
\eeq
Let us compute $\langle\hat \phi({\bf x},t)\hat \phi({\bf x}',t')\rangle$ in the large $n,\langle n
\rangle$ limit, corresponding to the regime in which can think of $\phi$ as a classical field:
\beq\label{eq:phiphi}
\begin{split}
&\langle\hat \phi^-({\bf x},t)\hat \phi^+({\bf x}',t')\rangle
=\sum_l \frac{1}{2\omega_l V} \langle n_l \rangle e^{-i {\bf k}_l\cdot({\bf x-x}')+i\omega_i (t-t')}\\
&\xrightarrow{\text{continuum}}  \int \frac{d^3 k}{(2\pi)^3} \frac{1}{2\omega_{\bf k}} \langle n_{\bf k} \rangle e^{-i {\bf k}\cdot{\bf \Delta  x}+i\omega_{\bf k}\delta t} \\
&\xrightarrow{\text{non-rel}} \frac{m^2e^{im\delta t}}2\int  \frac{d^3 v}{(2\pi)^3} \langle n_{\bf v} \rangle e^{-i m{\bf v}\cdot{\bf \Delta x}+\frac i 2 mv^2 \delta t}~.
\end{split}
\eeq
Here we defined $\bf \Delta x\equiv x-x'$, $\delta t=t-t'$. To find the normalisation of the field $\phi_0$, we can compute the its energy density
\beq
\begin{split}
\rho_\phi=&\frac{\langle (\dot\phi)^2+(\nabla\phi)^2+m^2\phi^2\rangle}2 
\simeq  m^4\int  \frac{d^3 v}{(2\pi)^3} \langle n_{\bf v} \rangle \\
\simeq& \frac12 m^2\phi_0^2\int  d^3 v\frac{2\langle n_{\bf v} \rangle}{(2\pi)^3}\frac{m^2}{\phi_0^2}
\equiv \frac12 m^2\phi_0^2\int  d^3 vf({\bf v}),
\end{split}
\eeq
where we traded $\langle n_{\bf v}\rangle$ with the normalized velocity distribution $f({\bf v})$ via
\beq
\langle n_{\bf v}\rangle \approx \frac{\phi_0^2}{2m^2}(2\pi)^3 f({\bf v})~.
\eeq
Therefore, \cref{eq:phiphi} can be rewritten as
\beq
\begin{split}
\langle\hat \phi^-({\bf x},t)\hat \phi^+({\bf x}',t')\rangle
\approx \frac{\phi_0^2}4 e^{im\delta t}\int  d^3 v f({\bf v}) e^{-i m{\bf v}\cdot{\bf\Delta x}+\frac i2mv^2 \delta t}.
\end{split}
\eeq

Assuming for simplicity that the dark matter velocity follows a Gaussian distribution \cref{eq:gaussian},
we can compute the first order degree of coherence for the dark matter field:
\begin{widetext}
\beq
\begin{split}
g^{(1)}({\bf x},t;{\bf x'},t')
\equiv &\frac{\langle\hat \phi^-({\bf x},t)\hat \phi^+({\bf x'},t')\rangle}{\sqrt{\langle\hat \phi^-({\bf x},t)\hat \phi^+({\bf x},t)\rangle \langle\hat \phi^-({\bf x'},t')\hat \phi^+({\bf x'},t')\rangle}}
=  e^{im\delta t}\int \frac { d^3 v }{\pi^{3/2}v_0^3}e^{-\frac{({\bf v+v_{\odot}})^2}{v_0^2}-i m{\bf v}\cdot{\bf \Delta x}+i\frac{mv^2}2 \delta t}\\
=& \frac{e^{im\delta t}}{(1+\xi^2)^{3/4}} e^{-i\frac{\xi \left(\frac{v_\odot^2}{v_0^2}-\zeta^2\right)-m{\bf v}_\odot\cdot{\bf \Delta x}}{1+\xi^2}-i\frac32\tan^{-1}\xi } e^{-\frac{(\xi v_\odot+\zeta v_0)^2}{v_0^2(1+\xi^2)}}
\end{split}
\eeq
\end{widetext}
where 
\beq
\xi\equiv \frac12 mv_0^2(t-t'),~\zeta\equiv \frac12 mv_0|{\bf x-x}'|.
\eeq

For $v_{\odot}\sim v_0$, the norm of $g^{(1)}|_{\bf x=x'}$ falls to $e^{-1}$ when $\xi\sim 1$. Thus $\tau_a\equiv\frac2{mv_0^2}$ is the coherence time for axion dark matter field. 
To get coherence length, consider 
\beq
g^{(1)}|_{t=t'}
= e^{-\zeta^2}e^{im v_{\odot} \cdot ({\bf x-x}')} 
\eeq
Therefore, $\frac 2{mv_0}$ can be identified as the axion field coherence length $\ldb$. 

\subsection{Signal Power in a Cavity}

$\langle  \hat \phi^-({\bf x}_1,t_1) \hat \phi^+  ({\bf x}_2,t_2)\rangle$ directly controls the outcome of a haloscope experiment, since the power spectral density, defined as $S_{\phi} (\omega)  \equiv \frac1T\langle \tilde \phi ({\bf x}_1,\omega)  \tilde \phi ^*({\bf x}_2,\omega)\rangle$, can be written as 
\begin{widetext}
\beq\label{eq:psd}
\begin{split}
S_{\phi} (\omega) &
= \frac{\phi_0^2} {4 T} \int dt_1 dt_2 e^{-i\omega (t_1-t_2)} \left(g^{(1)}({\bf x}_1,t_1;{\bf x}_2,t_2)+h.c.\right)
\approx  \frac{\phi_0^2} {4} \frac{\sin^2\left( \frac{(\omega-m) T}2\right)}{\left(\frac{(\omega-m)}2\right)^2T}e^{-\frac{|{\bf x}_1-{\bf x}_2|^2}{\ldb^2}}+(\omega\to-\omega)\\
&\xrightarrow{T\to\infty} \frac{\phi_0^2}2 \pi (\delta (\omega-m)+\delta (\omega+m))e^{-\frac{|{\bf x}_1-{\bf x}_2|^2}{\ldb^2}}
\end{split}
\eeq 
\end{widetext}
where we kept the leading order terms in both the oscillatory and non-oscillatory pieces of $g^{(1)}$ only.

Recall that to get the dark matter signal field in a cavity based experiment, one needs to solve \cite{Gao:2020anb} 
\begin{widetext}
\begin{equation}
\sum_n\left(\omega^2-\omega_n^2-i\frac{\omega\omega_n}{Q_n}\right){\bf E}_n({\bf x})\tilde e_n(\omega)\\
=\int dt e^{-i\omega t} \gag\partial_t({\bf B}_0\dot \phi({\bf x},t))
=-\omega^2 \gag {\bf B}_0 \tilde \phi({\bf x},\omega).
\end{equation}
\end{widetext}
Let the signal field be ${\bf E}_1$ and writing ${\bf B}_0=B_0\hat n$, we have
\beq
\sqrt{\int_V |{\bf E}_1|^2}\tilde e_1(\omega)
\\=-\frac{\gag B_0\omega^2}{\omega^2-\omega_1^2-i\frac{\omega\omega_1}{Q_1}}  \frac{\int _V {\bf E}^*_1 \cdot \hat {n}\tilde \phi(\omega)}{\sqrt{\int_V |{\bf E}_1|^2}}.
\eeq 

To read out the signal, we necessarily introduce additional losses in addition to the intrinsic loss. For now, let $Q_1^{-1}\to Q_{\rm eff}^{-1}=Q_0^{-1}+Q_{\rm e}^{-1}$, where $Q_0$ and $Q_{\rm e}$ are the intrinsic and extrinsic quality factors respectively. Now the signal power after readout is 
$P_{\rm sig}= \frac{\omega_1}{Q_{\rm e}}\int _V |{\bf E}_1|^2\langle e_1(t)e_1(t)\rangle$. Using $\langle \tilde \phi ({\bf x}_1,\omega)  \tilde \phi ^*({\bf x}_2,\omega')\rangle= S_{\phi} (\omega) 2\pi\delta(\omega-\omega')$, and $Q_{\rm e}=Q_0/\beta$, we find
\begin{widetext}
\beq\label{eq:psig}
P_{\rm sig}
=\frac{\omega_1\gag^2B_0^2\beta}{Q_0}\int \frac{d\omega}{2\pi}
\frac{\omega^4 }{(\omega^2-\omega_1^2)^2+\left(\frac{\omega\omega_1}{Q_0}(1+\beta)\right)^2}
\times \frac{\int _V \int_{V'} {\bf E}_1({\bf x'}) \cdot \hat n {\bf E}^*_1({\bf x}) \cdot \hat n S_{\phi} (\omega) }{ \int_V {\bf E}^*_1\cdot {\bf E}_1}
\eeq
\end{widetext}
For a homogeneous and monochromatic axion field $\frac{\sqrt{2\rho}}{m^2}\cos(mt)$, $S_\phi=\frac{\pi \rho}{m^2}(\delta(m-\omega)+\delta(m+\omega))$, and the above expression signal power gives the familiar result. Taking into account DM coherence over the detector by instead using $S_\phi$ from \cref{eq:psd}, \cref{eq:psig} yields on resonance $m=\omega_1$
\beq
\begin{split}
P_{\rm sig}
=&\frac{Q_0\gag^2B_0^2\rho V}{\omega_1}\frac{\beta}{(1+\beta)^2} \\&\times \frac{\int_{V} \int_{V'}  {\bf E}_1({\bf x'}) \cdot \hat n {\bf E}^*_1({\bf x}) \cdot \hat n e^{-\frac{|{\bf x}-{\bf x}'|^2}{\ldb^2}}}{V \int_V {\bf E}^*_1\cdot {\bf E}_1}.
\end{split}
\eeq
This gives precisely \cref{eq:generic_signal_power,eq:eta} for $\beta=1$.
Note that if the detector size $\sim |{\bf x}-{\bf x}'|\ll \ldb$, we recover the usual result. 

\section{Periodic Photonic Structures} 
\label{ap:periodic_structures}
Periodic photonic structures, often referred to as photonic crystals, are designed to control the flow of light in a manner analogous to how semiconductors manipulate the flow of electrons. These structures are characterized by their unique, regularly spaced arrangements of dielectric materials, which create a photonic bandgap – a range of frequencies where light cannot propagate. This control over the behavior of photons has given rise to a wide array of applications, from enhancing the efficiency of lasers and light-emitting diodes to enabling novel, compact optical devices, such as waveguides and filters. 
In this work we consider optical resonators made from simple periodic structures, such as 1D periodically grooved waveguide~\cite{Wang:17}, and 2D photonic crystal slab~\cite{PhysRevLett.109.067401}. The discrete translation invariance of the dielectric dictates that modes of the source-free Maxwell equations, i.e., without axion or dark photon DM, take the Bloch form:
\beq
\vecbf{E_\vecbf{K}} = \vecbf{u}_{\vecbf{K}}(\vecbf{r}) e^{\pm i \vecbf{K}\cdot \vecbf{r}}~,\quad  \vecbf{u}_{\vecbf{K}}(\vecbf{r})= \vecbf{u}_{\vecbf{K}}(\vecbf{r}+\vecbf{R})
\label{eq:Bloch_field_parametrization}
\eeq
where $\vecbf{K}(\omega)$ is the Bloch wavevector (in the first Brillouin zone) and $\vecbf{R}$ is any lattice vector.
Solving the source-free Maxwell equations fixes the dispersion relation $\omega = \omega(\vecbf{K})$. 
In a photonic resonator comprised of $N_u$ unit cells, the overlap factor can be simplified as
\beq
|\eta|^2  = |\eta_u|^2 \frac{1}{N_u^2}\sum_{i,j}  e^{-i \vecbf{K}\cdot (\vecbf{R}_i -\vecbf{R}_j)}~,
\label{eq:resonator_overlap_factor_coherent_dm}
\eeq
where
\beq\label{eq:eta_u}
\eta_u = \frac{\frac{1}{V_u}\int_u d^3 r e^{-i \vecbf{K}\cdot \vecbf{r}} \vecbf{u}_\vecbf{K}(\vecbf{r})\cdot \hat{\vecbf{n}}}{\sqrt{\frac{1}{V_u}\int_u d^3 r \varepsilon(\vecbf{r})|\vecbf{u}_{\vecbf{K}}(\vecbf{r})|^2}}~,
\eeq
is the single cell overlap, with $V_u$ the volume of each unit cell.
From \cref{eq:resonator_overlap_factor_coherent_dm} we see that the overlap factor is non-zero if $\vecbf{K} = 0$ and $\eta_u \neq 0$.

\subsection{1D-Periodic Resonators}
\label{sec:1d_resonators}
First consider a structure that is periodic along a single direction, $\hat{\vecbf{z}}$, i.e. $\varepsilon(x,y,z)=\varepsilon(x,y,z+\Lambda)$, where $\Lambda$ is the length of the unit cell. Assuming that the DM is coherent over the entire structure (i.e., we can neglect variations in DM field over the device) the overlap factor, \cref{eq:resonator_overlap_factor_coherent_dm}, simplifies to
\beq
|\eta|^2=|\eta_u p_{N_u}(K\Lambda)|^2,~p_{N_u}(\theta) \equiv 
\frac{1-e^{-i N_u\theta}}{N_u(1-e^{-i\theta})}\label{eq:pN}.
\eeq
In the limit of $N_u\gg 1$, $|p_N(K \Lambda)|^2$ is \emph{primarily} peaked at $K=0$ as expected from approximate momentum conservation. 

One type of 1D-periodic resonator is a microring (see the insert in \cref{fig:setup}), which we can form from $N_u$ unit cells by enforcing an additional periodicity condition $\varepsilon(x,y,z)=\varepsilon(x,y,z+L)$ where the ring circumference is $L = N_u\Lambda$. The field modes in \cref{eq:Bloch_field_parametrization} must be similarly periodic, which selects a discrete subset $K = 2\pi n / L$ for integer $n$ which can be resonantly enhanced. The DM-coupled $K=0$ modes automatically satisfy this condition.

A non-zero DM signal requires $\eta_u \neq 0$. Below we show that this is indeed possible in a 1D-periodic dielectric waveguide with transverse confinement.\footnote{The case without transverse mode confinement corresponds to the wide-area dielectric stack studied in~\cite{Baryakhtar:2018doz}.} An analytic treatment is possible only for highly-symmetric geometries so we study a cylindrical periodically-varying waveguide. Other geometries and resonator form-factors generally must be investigated numerically. 

\subsubsection*{Cylindrical Fibre Bragg Grating} 
By introducing large permittivity modulations periodically along the propagation direction, one can achieve perfect confinement of light even in the presence of radiating modes in the free space (see, e.g., Refs.~\cite{PhysRevB.74.115125, PhysRevA.96.013841,PhysRevLett.122.187402,gao2019bound,PhysRevApplied.15.034041}). 
For simplicity, we consider a step index circular waveguide consisting of a core of radius $R$ and a periodic $\varepsilon(z)$ with periodicity $\Lambda$.  The cladding has a  constant refractive index $n_o$ and a radius much bigger than $R$. This configuration is also known as a fibre Bragg grating. 
To accurately describe the leaky fiber
modes, below we employ the Fourier modal method~\cite{Armaroli:08}, which handles the radiating fields analytically. 
Using the cylindrical coordinates $(r,z,\phi)$, let ${\bf E}\sim (E_r,E_z,E_\phi) e^{im\phi-i\omega t}$, ${\bf H}\sim (H_r,H_z,H_\phi) e^{im\phi-i\omega t}$, where $m$ is an integer representing the angular momentum in the plane. The $z-$components of the wave equations are given by
\beq
\hat L_m\left(
\begin{array}{c}
  E_z   \\
    H_z  
\end{array}\right)
= \left(\begin{array}{l}
   -(\partial_z\frac{\varepsilon'}{\varepsilon}+\frac{\varepsilon'}{\varepsilon}\partial_z+\omega^2\varepsilon)E_z\\
     -\omega^2\varepsilon H_z
\end{array}\right)
\eeq
where $\hat L_m\equiv (\frac1r\partial_rr\partial_r-\frac{m^2}{r^2}+\partial_z^2)$. 
The solutions take the following form:
\beq\left(
\begin{array}{c}
\sqrt{\varepsilon} E_z\\
H_z
\end{array}\right)
=e^{iKz}\sum_ne^{i2n\pi z/\Lambda }\left(
\begin{array}{c}
\psi_n^+(r)\\
\psi_n^-(r)
\end{array}\right)
\eeq
where $+(-)$ represents TM (TE) modes, and $n$ are integers. It can be shown that 
\beq
\psi_n^{\sigma\in\{+,-\}}\sim \left\{
\begin{array}{cc}
\sum_l P^\sigma_{nl} J_m(\lambda^\sigma_l r) ,& r<R \\
H^{(1)}_m(\alpha_n r) ,& r>R\\
\end{array}\right .
\eeq
where $J_m$ and $H_m^{(1)}$ are Bessel functions and Hankel functions of the first kind, and  
\beq
\alpha_n=\sqrt{\omega^2n_o^2-(K+2n\pi/\Lambda)^2}~.
\eeq 
For confined modes, $\alpha_n$ needs to be imaginary. Thus, when $K=0$, the $H^{(1)}_m(\alpha_0 r)$ mode is necessarily leaky. The periodic $\varepsilon(z)$ mixes different radial modes via the mixing matrix $P_{nl}$. More details of $\psi_n^\sigma$ can be found in, for instance, \cite{PhysRevB.74.115125}.

The $\phi$ components of the fields can be inferred from the $z-$components and will take the following form: 
\beq\left(
\begin{array}{c}
\frac{H_\phi}{\sqrt\varepsilon}\\
E_\phi
\end{array}\right)
=e^{iKz}\sum_ne^{i2n\pi z/\Lambda}\left(
\begin{array}{c}
\chi_n^+(r)\\
\chi_n^-(r)
\end{array}\right),
\eeq
where $\chi_n^\pm$ are related to $\psi_n^\pm$ by Maxwell's equations. The solutions are obtained by demanding the continuity of $E_z,E_\phi,H_z,H_\phi$ at $r=R$. 
After solving the system, $E_r$ can be obtained via
\begin{align}
 E_r=\frac{-m}{\varepsilon \omega r} H_z+\frac{i}{\omega\varepsilon }\partial_z H_\phi~.
\end{align}
For an arbitrary $R$, $\omega$ is generally complex which determines the intrinsic quality factor of the mode: 
\beq
Q=-\frac{\re\omega}{2\im \omega}.
\eeq
One can fine tune the radius of the core $R_0$ to find a corresponding real solution $\omega_0$, which will be completely confined without considering the additional  loss due to bending or surface roughness.

For axion dark matter searches, an external magnetic field will be imposed. If the magnetic field points in the transverse direction of the Bragg grated fibre, say $y-$direction, the modes of interest must break rotational symmetry, i.e. $m>0$, which will be a hybrid mode of TE and TM polarization. Consider the case of $m=1, K=0$. We can estimate $\eta_u$ by considering the integral $V^{-1} \int_V E_y$ which in terms of cylindrical coordinates is given by
\beq\label{eq:etacylinder}
\int_0^{2\pi} \frac{d\phi}{2\pi} \int_0^\Lambda\frac{dz}{\Lambda}\int_0^{R^2}\frac{dr^2}{R^2} e^{i\phi} \left(E_r\sin\phi+E_\phi \cos\phi\right)~.
\eeq 
Given the form of $E_\phi$, integrating over $z$ will pick out the $n=0$ mode. However, $n=0$ mode is generally leaky and thus should be highly suppressed for a high $Q$ solution. Similarly, the form of $E_r$ implies that the $\frac{i}{\omega\varepsilon }\partial_z H_\phi$ part should dominantly contribute to the overlap integral, which becomes
\beq
\frac1 V\int_V E_y 
\approx \frac12\int_0^\Lambda\frac{dz}{\Lambda}\int \frac{dr^2}{R^2} \frac{\partial_z\varepsilon^{-1}}{\omega} H_\phi~.
\eeq 

For dark photon dark matter searches, $m=1,K=0$ mode would overlap with a transverse dark photon field, whereas $m=0,K=0$ mode would overlap with a longitudinal dark photon field. For the latter,  $\eta_u$ is determined by $V^{-1}\int_V E_z$. 

Therefore, both the axion and dark photon dark matter dominantly couple to the TM polarization of the zero-$K$ modes. As an example, consider  
\beq\label{eq:epsilon}
\varepsilon(z)=\left\{
\begin{array}{cc}
\varepsilon_1,& 0<z<a\\
\varepsilon_2, & a<z<\Lambda
\end{array}
\right . .
\eeq
This yields an overlap factor which is approximately
\beq
\eta_u\sim \left\{
\begin{array}{l c}
\sqrt2\frac{ (\varepsilon_1-\varepsilon_2)} { (\varepsilon_1+\varepsilon_2)^{3/2}}, & m=1\\
\frac{\varepsilon_1^{1/2}-\varepsilon_2^{1/2}}{\sqrt{2\pi\varepsilon_1\varepsilon_2}}\sin\left(\frac{2a\pi}\Lambda\right), &m=0
\end{array}\right .
\eeq
where we assume that integrating the radial part yields a factor $\lesssim1$. 

\begin{figure}
    \centering    \includegraphics[width=0.4\textwidth]{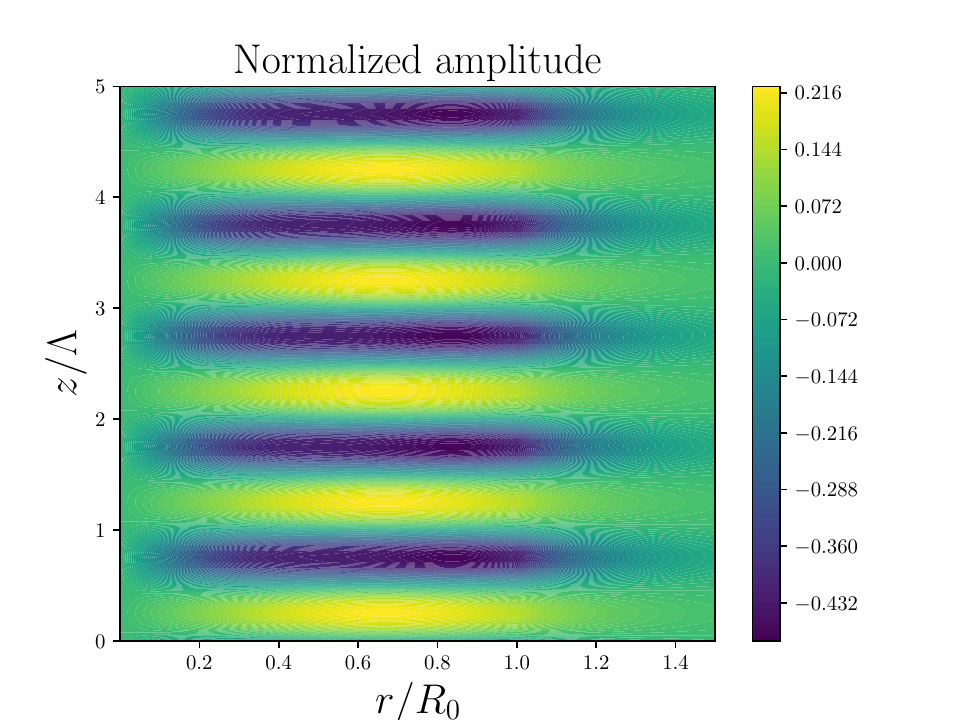}
    \caption{The normalized amplitude $\left(\sqrt{V^-1\int_V \varepsilon|{\bf E}|^2}\right)^{-1}\int_0^{2\pi}\frac{d\phi}{2\pi} E_y(r,z) \times \frac{r}{R_0}$ for a fine-tuned  $m=1,K=0$ confined mode in a cylindrical fiber Bragg grating is shown. This mode has a non-zero overlap factor $\eta$ and therefore couples to the DM source. We took a a periodic relative permittivity $\varepsilon=
\frac{1}{2} (\varepsilon_1-\varepsilon_2)  \left[1+\sin (2 \pi z/\Lambda)\right]+\varepsilon_2$ with $\varepsilon_2=1,\varepsilon_1=5$ in the core and $n_o=1$ in the cladding.}
    \label{fig:Ey_bgf}
\end{figure}

To numerically verify this, we consider a toy example with a periodic permittivity in the core given by $\frac{1}{2} (\varepsilon_1-\varepsilon_2)  \left[1+\sin (2 \pi z/\Lambda)\right]+\varepsilon_2$ with $\varepsilon_2=1,\varepsilon_1=5$, and $n_o=1$ in the cladding. We work in the basis of $n=0,\pm1$ modes only, and fine tune the radius to obtain a confined mode with $m=1,K=0$, which has $\tilde\omega_0\equiv \omega_0/(2\pi/\Lambda)=0.649038 , R_0=0.942518/(2\pi /\Lambda)$ (c.f. Figure~\ref{fig:Ey_bgf}). 
Up to an overall normalization, the solutions are given by
\beq
\sqrt{\varepsilon}E_z=2i\cos\left(\frac{2\pi z}\Lambda\right)\left\{
\begin{array}{cc}
 \frac{J_1(\lambda_1^+r)}{J_1(\lambda_1^+R_0)},   & r<R_0\\
\frac{H^{(1)}_1(\alpha_1r)}{H^{(1)}_1(\alpha_1 R_0)},   &r>R_0
\end{array}\right .
\eeq
with $\lambda_1^+=0.0918145(2\pi/\Lambda)$, and
\begin{widetext}
\begin{align}
H_z(r<R_0)=&
-\frac{i}{2\tilde\omega_0^2}\left[\frac{ g_2^-J_1(\lambda_2^-r)}{J_1(\lambda_2^-R_0)}-\frac{g_3^-J_1(\lambda_3^-r)}{J_1(\lambda_3^-R_0)}\right]
+\frac{i\sqrt{1+8\tilde\omega_0^4}}{2\tilde\omega_0^2}\left[\frac{ g_2^-J_1(\lambda_2^-r)}{J_1(\lambda_2^-R_0)}+\frac{g_3^-J_1(\lambda_3^-r)}{J_1(\lambda_3^-R_0)}\right]\\
&-2i\sin\left(\frac{2\pi z}\Lambda\right)\left[
\frac{ g_2^-J_1(\lambda_2^-r)}{J_1(\lambda_2^-R_0)}+\frac{g_3^-J_1(\lambda_3^-r)}{J_1(\lambda_3^-R_0)}\right],\nonumber\\
H_z(r>R_0)=&-2i\sin\left(\frac{2\pi z}\Lambda\right)
\frac{(g_2^-+g_3^-)H^{(1)}_1(\alpha_1r)}{H^{(1)}_1(\alpha_1 R_0)}
\end{align}
\end{widetext}
with $\lambda_2^-=i0.11834(2\pi/\Lambda), \lambda_3^-=1.24157(2\pi/\Lambda), g_2^-=0.592891,g_3^-=0.128881$. Note that an accidental cancellation happens for the TE polarization such that $H_1^{(1)}(\alpha_0 r)$ vanishes.
Using \cref{eq:etacylinder}, this solution gives an overlap factor of $|\eta_u|\approx 0.1$.

\subsection{2D-Periodic Resonators}
\label{sec:2d_resonators}
Dielectric ring resonators suffer from ``inevitable'' losses of photons due to their curvature and due to the sharp interfaces between domains of different indices of refraction. Both effects limit the maximum quality factor that can be attained in practice. Despite this, current state-of-the art resonators reach $Q\sim 10^6$ in periodically-poled (see Table 2 in Ref.~\cite{Zhu:21}) and $\sim10^4$~(see, e.g., Refs.~\cite{Flueckiger:16,grated_ring_high_Q}) in periodically-grooved resonators \emph{for circularly propagating modes}. As shown above, DM couples to standing wave-like modes with $\vecbf{K} \approx 0$, for which $Q$ values are likely lower since they lie above the light-line $\omega = |\vecbf{K}|/n_{air}$ (i.e., these states can couple to continuum radiation). Photonic structures that are periodic in 2D, however, are known to support Bound States in Continuum (BIC) which are extremely long-lived modes with $|\vecbf{K}| \ll \omega$~\cite{hsu2016bound}. We expect DM to couple most strongly to these $\vecbf{K} \approx 0$ modes.

\begin{figure}
    \centering
    \includegraphics[width=0.4\textwidth]{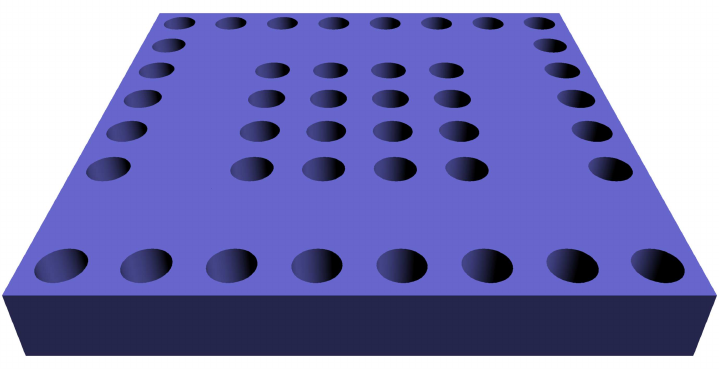}
    \caption{A sketch of a photonic slab resonator (the periodic hole pattern in the middle) and the surrounding waveguides (gaps in the periodic hole pattern). For ease of illustration the resonator is drawn with only 16 unit cells. Wave-like DM can excite a mode in the resonator which sources fields in the waveguides.}
    \label{fig:photonic_slab}
\end{figure}

The large $Q$ values possible in BICs arise due to accidental cancellations between different radiation modes, or, due to a high degree of symmetry in the dielectric arrangement. In this section we explore the latter possibility by considering a dielectric slab with a square lattice of holes -- see Fig.~\ref{fig:photonic_slab}. The discrete rotational symmetry of the slab enables a categorization of all modes in terms of its irreducible representations (irreps) of the symmetry group; BICs and radiation modes fall into different irreps of the group, such that they formally have 0 overlap leading to a (formally) infinite quality factor~\cite{PhysRevB.65.235112}. In practice, the quality factor is controlled by the finite size of the slab, and manufacturing imperfections (e.g., variations in the size and positions of the holes), leading to realistic quality factors of $10^4$~\cite{PhysRevLett.109.067401}. Even higher quality factors $\sim 5\times 10^5$ have been engineered by tuning slab geometry to combine multiple BICs~\cite{jin2019topologically}.

The device sketched in Fig.~\ref{fig:photonic_slab} consists of a square lattice of holes in the center serving as the resonator. The gaps (``line defects'') surrounding the finite lattice act as highly effective (low loss) waveguides~\cite{lonvcar2000design}. DM excites a standing wave mode in the resonator which leaks out into the gaps which guide the signal photons onto a detector. 

The formalism described in \cref{sec:electrodynamics_with_DM} and the beginning of \cref{ap:periodic_structures} applies to 2D-periodic structures. In fact, we can now slightly extend the previous discussion and allow for the resonator to be larger than a DM coherence length. This means that while individual unit cells still have a size much smaller than $\ldb$, their spatial separation can be much larger than $\ldb$. In this situation the overlap factor can be written as 
\beq
|\eta|^2  = |\eta_u|^2 \frac{1}{N_u^2}\sum_{i,j}  e^{-i \vecbf{K}\cdot (\vecbf{R}_i -\vecbf{R}_j)}e^{-(\vecbf{R}_i- \vecbf{R}_{j})^2/\ldb^2}~,
\label{eq:resonator_overlap_factor_incoherent_dm}
\eeq
This modification means that at a fixed $\vecbf{K}$ (correspondingly fixed $\omega$), only those unit cells with $|\vecbf{R}_i- \vecbf{R}_{j}|/\ldb < 1$ contribute significantly to the overlap. By approximating the double sum in ~\cref{eq:resonator_overlap_factor_incoherent_dm} by a double (two-dimensional) integral, it is straightforward to show that the result scales as $\ldb^2/(N_u A_u)$, such that the full signal power loses total volume scaling. Instead, for a fixed signal mode with Bloch momentum $\vecbf{K}$, the volume factor $V = N_u A_u h$ in \cref{eq:generic_signal_power} is effectively replaced by $\ldb^2 h$, i.e. the signal power appears to come from a single ``coherence volume'' of the detector. Extensive scaling of the signal power is restored if the detector is sensitive to a range of $|\vecbf{K}|\sim 1/\ldb$ values that the DM couples to. The result is that the signal power scales with total detector volume and \cref{eq:generic_signal_power} holds after summing over $\vecbf{K}$.\footnote{This situation is completely analogous to the dielectric stack which has a much larger area than $\ldb^2$~\cite{Baryakhtar:2018doz}. As a result, the stack emits photons with a range of transverse momenta $\vecbf{k}_\perp$ as dictated the DM velocity distribution. The summation over these different modes is accomplished by the lens which redirects these photons onto a single detector. Only after this step the signal power proportional to the total area of the stack.} 

Patterned slabs appear to have several useful features for DM detection. First, the effective index contrast between different regions within a unit cell is large, allowing us to maximize the overlap factor. Second, slabs feature a larger space-filling fraction on a substrate wafer, leading to a larger instrumented volume compared to, e.g., ring resonators on a same-sized wafer.  Finally, the patterned slabs can have large areas $\sim \mathrm{cm}^2$~\cite{PhysRevLett.109.067401} at a single lattice spacing; alternatively, multiple slabs with different lattice spacings can be placed on a single line-defect waveguide. The latter set-up is analogous to placing multiple ring resonators on a single bus. 
It would be interesting to study whether 3D-periodic structures have further advantages. We leave the exploration of DM signals in these systems to future work.

\subsubsection*{2D Example: Square Lattice of Holes}
In this section we discuss some details of the square lattice photonic slab as a DM detector. Our primary aim is to show that this structure supports low-loss modes that couple to DM. The low loss condition implies that the signal mode does not couple to radiation modes, while a non-zero coupling to DM requires a non-zero overlap integral. We can gain insight into both requirements by considering symmetries of the slab.

The square lattice is symmetric under reflections about $z$ plane and the point group $C_{4v}$~\cite{alma991011485189705164}, which consists of $\pi/2$ rotations, reflections through the $xz$, $yz$ planes and through two other diagonal planes. In the idealized case of an infinite slab with perfect holes, the only losses are due to radiation above or below the slab, where the modes must reduce to plane waves propagating in $\pm z$. Since there are two polarizations, these radiation modes must correspond to two-dimensional irreps of $C_{4v}$. Therefore modes falling into one-dimensional (singlet) irreps (called $A_1$ and $B_1$) of $C_{4v}$ cannot radiate to infinity. These are the low-loss modes we are after. 

\begin{figure}
    \centering
\includegraphics[width=0.47\textwidth]{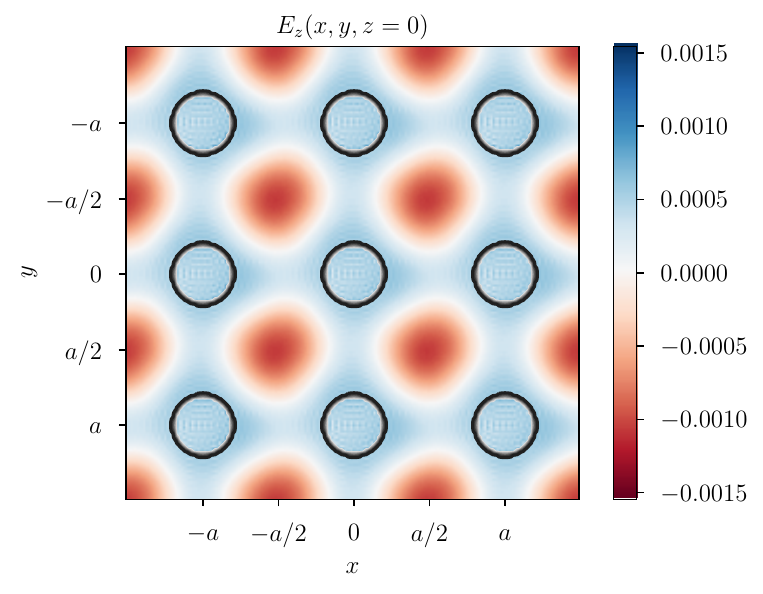}
    \caption{Example electric field mode (in the $z=0$ plane) of the square lattice of air holes that is both long-lived (symmetry-protected from coupling to radiation modes) and has a non-vanishing overlap with DM. The axes are given in units of the lattice constant $a$. The fields were found using \texttt{MPB} for slab height $0.5a$, hole radius $0.2a$ and slab dielectric constant $\varepsilon = 12$. The electric field amplitude is normalized to such that $\int d^3 r \varepsilon |\vecbf{E}|^2 = 1$ over one unit cell. The black circles are the air holes in the dielectric slab. Only three periods of the photonic structure are shown in each direction. }
    \label{fig:square_lattice_signal_mode}
\end{figure}

Now we can apply the condition of a non-zero overlap. This eliminates modes which are not symmetric under all reflections, leaving only $A_1$ irreps. Typically the mode with the largest overlap will have the fewest possible nodes, so we conclude that the lowest-energy $A_1$ will have the largest overlap factor. Such a mode is shown in Fig.~\ref{fig:square_lattice_signal_mode} for slab height $0.5a$, hole radius $0.2a$ and slab dielectric constant $\varepsilon = 12$ where $a$ is the length of the unit cell. This mode was found using \texttt{MPB}~\cite{johnson2001block}.\footnote{\url{https://github.com/NanoComp/mpb}} Numerically we find $\eta_u \approx 10^{-2}$. We leave the optimization of the overlap factor through variations in geometry to future work. It will also be interesting to explore other lattices. 

\section{Signal Rates}\label{sec:signal_rate}
\subsection{Single Resonator}
Assuming that the resonator is a microring, let us estimate the signal rate from a single resonator with the target frequency $\omega_R$. The largest 
overlap factors are expected to arise from the EM modes with fewest nodes, which means that the length of a single unit cell is approximately $2\pi/\omega_R$, leading to a resonator circumference of
\beq\label{eq:L}
L = 2\pi/\omega_R \times N_u~,
\eeq
where $N_u$ is the number of unit cells. The cross-sectional area of the resonator is typically 
\beq\label{eq:A}
A= t_R\times 2\pi/\omega_R~,
\eeq
where $t_R$ is the thickness of the resonator.
Using \cref{eq:P1_res} with \cref{eq:dm_avg_current}, the signal rate from one critically-coupled ($\tau_\mathrm{e}^{-1} = \tau^{-1}/2$) resonator on resonance is
\beq\label{eq:sig}
\Gamma_{1}
\sim \frac{10^{-11}}{\rm sec} 
\left\{
\begin{array}{c}
0.2 \left(\frac{\rm 0.25 eV}{\omega_R}\right)^4\left(\frac{g_{a\gamma}{\rm GeV}}{10^{-10}}\right)^2 \left(\frac{B}{9.4{\rm T}}\right)^2\\
10^{9}\left(\frac{\rm 0.25 eV}{\omega_R}\right)^2\left(\frac{\chi}{10^{-10}}\right)^2
\end{array}\right .
\eeq 
where we take $Q=5000$, $|\eta_u|^2\sim 0.1$, $N_u=100$, $t_R=10\;\mu$m.
Here the first line corresponds to axion DM and the second to dark photon DM.
For both models, the coupling ($\gag$ or $\chi$) has been normalized to a value near the current experimental bounds~\cite{AxionLimits,Caputo:2021eaa} which arise from modifications to stellar evolution in the presence of new, weakly coupled particles. 
Note that the signal rate for dark photon DM is significantly higher than that of axion DM in this mass range. 
Even over one year, the number of events in a single resonator from axion DM at the currently allowed couplings is significantly less than $1$. 
To increase the signal rate, signals from a large number of resonators need to be combined, which leverages the scaling advantages of integrated photonics.

\subsection{Multi-Resonator Setup}\label{sec:multi_sig_rate}
We can simultaneously readout many resonators to improve the scaling in two distinct ways. 

First, the output of multiple resonators of the same frequency may be summed to increase the search sensitivity at a specific $\mdm$. To overcome the limited DM coherence over a large number of resonators, a multi-mode readout scheme, such as spatial power combining, should be employed, 
as discussed in 
\cref{ap:SI-combining}.

Second, which is also the setup employed in deriving the projected sensitivities in this work, is described below.
The key is to realize that an arbitrary number of resonators of different frequencies can be coupled to a single bus, as there are no interference effects between signals of differing frequency. See \cref{fig:width} for a study of this arrangement.
Thus, we may fabricate resonators of varying $\omega_R$ on the same wafer.
In practice, we could also scan $\mdm$ by tuning the resonators with applied heating and the electro-optic effect, but for simplicity here we will treat each $\omega_R$ as fixed.  

As depicted in \cref{fig:setup}, our proposed  experimental setup consists of a collection of wafers, each containing an array of resonators and readout busses.  
Each bus will couple to resonators of $N$ distinct frequencies, with busses on the same photonic chip directed onto a single photon counter.
Below we consider the sensitivity for some reasonable, benchmark parameters and we do not optimize here over all design choices, such as the frequency coverage per wafer, etc., as this ought to consider various factors beyond the scope of this work. 

For a wafer of diameter $D$ we estimate the total number of resonators that it can accommodate as follows. 
Let the radius of the smallest-frequency microring be $R = N_u/\omega_R$ and let all resonators be spaced by $\sim 2R$. 
The smallest frequency resonator has the largest size, so this ensures no spatial overlap of resonators.  Thus,
\eqa{
\la{eq:NR}
 N_\tx{W} \sim \pfrac{D \omega_R }{ 2 N_u }^2
}
resonators can be fitted onto a single wafer.
For a $15 \cm$ diameter wafer, with $N_u = 100$ and the smallest frequency of $\omega_R = 1 \,\eV$ or $2\pi/\omega_R = 1.24\, \micron$, this gives $\sim 10^6$ resonators. 

The fraction $\xi_\tx{act}$ of active volume, i.e.~total volume inside the resonators, is still small. 
Given that a single wafer has a volume of 
$V_{\rm wafer}=\pi D^2t_\tx{W}$ with $t_\tx{W}$ its thickness, in terms of the resonator count per wafer $N_\tx{W}$, the active fraction is found to be
\eqa{
\la{eq:xi-NW}
\xi_\tx{act} 
  = \frac{N_\tx{W} L A}{V_{\rm wafer}}\sim N_{\rm W} \left(\frac{2 \pi }{\omega_RD} \right)^2  \frac{N_u t_{\rm R}}{\pi t_\tx{W}}
,}
where ~\cref{eq:L,eq:A} are used. With the maximum resonator count \cref{eq:NR} we have 
\eqa{
\la{eq:xi}
\xi_\tx{act} 
&\sim 1\% 
\pfrac{100}{N_u} 
\pfrac{t_{\rm R}/t_\tx{W}}{0.1}~.
}

\subsection{Total Volume}
Despite large single-wafer enhancement factors, experiments seeking discover DM beyond existing constraints must instrument larger volumes by combining multiple wafers. 
Given a total volume $V_{\rm act}$ of the resonators from all wafers that cover $N$ different DM frequencies, the signal rate is enhanced by a factor corresponding to the total number of resonators per frequency,
\eqa{
    \Gamma_\tx{sig} \approx & \, 4 \cdot 10^8 \, \Gamma_1  \,
    \pfrac{100}{N}
    \pfrac{V_\tx{act}}{1000\; \cm^3} \times \nonumber \\ 
     \times & \pfrac{\omega_R}{0.25\; \eV}^2
    \pfrac{100}{N_u}
    \pfrac{10\; \micron}{t_{\rm R}}
}
where $\Gamma_1$ is given in \cref{eq:sig}.
The maximum possible active volume $V_{\rm act}$ in any axion search is controlled by $\xi_{\rm act}$ times the magnet bore volume. There is no such limitation for dark photon searches. 
In \cref{tab:magnets} we list several strong, large-bore magnets that either exist or are being considered. 
It is clear that larger bore sizes generally yield a bigger $B^2 V_{\rm act}$.
Given a fixed $B^2 V_{\rm act}$ that covers $N$ axion frequencies, the axion DM signal rate becomes
\beq
\begin{split}
&\Gamma_{\rm sig}(\omega_R)\sim \frac{8\times 10^{-4}}{\rm sec} 
\left(\frac{Q}{5000}\right)\left(\frac{|\eta|^2}{0.1}\right)\times\\
&\left(\frac{0.25\;\eV}{\omega_R}\right)^2
\left(\frac{g_{a\gamma}{\rm GeV}}{10^{-10}}\right)^2 \left(\frac{B^2V_{\rm act}}{67~\rm kJ}\right)\left(\frac{100}N\right).
\end{split}
\eeq

In this approach, the signal is limited by the volume of the large-bore magnet. 
To avoid this, one could use small Nd button magnets ($B < 1$ T) which source a large field only in the resonator volumes, allowing in principle the use of an arbitrarily large collection of resonators.  
It is also possible that a careful placement of such magnets can provide phase-matching to axion DM without the need for a periodic optical structure.  
To improve signal-to-noise, one could employ pulsed magnetic fields which could reach up to 100 T, such as those at U.S. National Magnet Laboratory, and would allow time-gating the detection signal to further reduce dark counts.
We leave these considerations for future work, and for simplicity focus here on a large-bore magnet as have been used in prior axion searches.  

\begin{table}[b]
    \centering
\begin{tabular}{cccc c}
    \hline
  $B$ (T) & Bore (mm) & $V_{\rm act}$ (cm$^3$) & $B^2V_{\rm act}\;({\rm kJ})$ & References  \\
    \hline
   40 & 34  &$9\times 10^{-2}$ &$0.11$ &\cite{10061544} \\
  21 & 123 & 1.18 &$0.40$&\cite{doi:10.1007/s13361-015-1182-2} \\ 
9.4 & 800 & 1000 &$67.3$  & \cite{thulborn2016quantitative}\\
 11.7 & 900 &1270 & $133 $ & \cite{Boulant2023}\\
  20\footnote{A future 20 Tesla MRI magnet} & 680  & 726 &$221 $&\cite{BUDINGER2018509} \\
    \hline
\end{tabular}
\caption{Realistic magnet parameters that can be employed in an axion search. All entries are MRI magnets except for the 40 Tesla magnet. $V_{\rm act}$ is the product of the physical volume of the bore and the active fraction $\xi_{\rm act}\sim 1\%$, defined in \cref{sec:signal_rate}. }
    \label{tab:magnets}
\end{table} 

\subsection{Backgrounds}\label{sec:bkg}
The coupling reach is enhanced if a larger active volume 
can be obtained. As depicted in Figure~\ref{fig:setup}, this requires a large number of buses and wafers, which in turn requires a large detector area to ensure a high detection efficiency. One can estimate the required detector area by
\beq
A_{\rm d}\approx \frac{V_{\rm act}/\xi_{\rm act}}{V_{\rm wafer}}\sqrt{N_{\rm W}}A \sim \frac{V_{\rm act}}{\sqrt{N_{\rm W}}L},
\eeq 
where Eq.~\eqref{eq:xi-NW} is used.
Thus, for a given $V_{\rm act}$ made up from wafers of diameter $D$, the typical detector area needed is
\eqa{
\label{eq:area}
  A_\tx{d} \approx
  20 \cm^2 \pfrac{V_\tx{act}}{ 1000 \cm^3}\pfrac{15\cm}{D}   
}
where \cref{eq:L,eq:NR} are used. Note that this is independent of frequency. 

Alternatively, one could photonic wire bond independent wafers together to avoid requiring a detector element for each wafer, and thus increase the signal-to-noise ratio.
The technical feasibility of this approach is left for future investigation.

We consider two example detector systems, Skipper charge-coupled device (CCD) and superconducting nanowires (SNSPDs)~\cite{Chou:2023hcc}. 
The frequency sensitivity range, typical areas and dark count rates ($\mathrm{DC}$s) of single photodetectors are listed in \Tab{tab:dc}. 
Over our full detector area in \cref{eq:area} we have a background timescale of 
\eqa{
\Gamma_\tx{bkg}^{-1} \approx 
\begin{cases} 
  13.3 
  \sec , & \omega < 1.12 \eV  \; \tx{(SNSPD)} \\ 
 112.5 \sec , &  \omega > 1.12 \eV  \; \tx{(CCD)}.\\
  \end{cases}
}

\section{Combining Signals}
\label{ap:SI-combining}
The optimal signal power from $N$ resonators is proportional to the total volume. Thus, $N$ identical resonators have collectively an optimal power of $N P_0$, where $P_0$ is the optimal power of a single resonator. 
One might have thought that by combining signals in-phase an optimal power $\propto N^2$ could be achieved, but this is incorrect. 
Conservation of energy implies that the power input to the resonator by the DM source, the power lost to dissipation, and the signal power drawn from the resonator obey
\eqa{
  P_\tx{source} = \Psig + P_\tx{loss}.
}
For a resonator of quality $Q$ driven by a fixed source current $\Jdm$, on resonance we have that the source and loss power are determined by the resonator mode $\vecbf{E}_p$ and mode amplitude $e_p$ as
\eqa{
  P_\tx{source} &= \frac{1}{2} \Re \lb  e_p^* \int d^3x \; \vecbf{E}_p^* \cdot \Jdm \rb \\
  P_\tx{loss} &= \frac{\omega |e_p|^2}{2Q} \int d^3x \; |\vecbf{E}_p|^2.
}
Since these two quantities scale with different powers of $e_p$ it follows that the signal power is bounded from above, 
\eqa{
    \Psig \leq \frac{Q}{8 \omega}  
    \lp \frac{|\int d^3x \; \vecbf{E}_p^* \cdot \Jdm|^2}
    {\int d^3x \; |\vecbf{E}_p|^2} \rp
    \equiv P_c
\label{eq:Pc}
}
which is saturated at a particular, critical value of the mode excitation $e_\tx{crit}$. 
This corresponds to the case in which the coupling rate of the resonator to the bus is matched to the loss rate of the resonator.
This is essentially the result in \cref{eq:generic_signal_power,eq:eta}, save that those expressions further consider the microphysics of $\Jdm$ generated by a stochastic DM field. Simultaneously reading out additional resonators does nothing to alter this bound, so long as the resonators are sufficiently separated that their presence does not alter the form of $\vecbf{E}_p$.  
Thus a system of $N$ identical resonators can produce a signal power of at most $N P_c$.  

\subsection{Collective Readout}
We now examine the requirements on a readout scheme to achieve the limit $N P_c$. 
Consider $N$ identical resonators with no mutual interactions and assume that the DM coherence length is larger than the resonator size, but not necessarily larger than the distance between resonators. 
The lack of interactions is a simplification.
The interactions are included in the next section.
Let each resonator have one mode $\vecbf{E}_p$ with frequency $\omega_p$ and assume the  modes of different resonators are disjoint in space.
Maxwell's equation may then be considered resonator-by-resonator, and the excitation $e_i$ of the $i^\tx{th}$ resonator obeys
\eqa{
    \lp \partial_t^2 + \frac{\omega_p}{Q} \partial_t + \omega_p^2 \rp e_i \, \vecbf{E}_p 
    = - e^{i \alpha_i} \partial_t \Jdm& \\ 
\Rightarrow \;\; e_i = 
   \frac{- i \omega \, e^{i \alpha_i}}{\omega_p^2 - \omega^2 + i \frac{\omega \omega_p}{Q}} 
    \lp \frac{\int_{V_i} d^3x \; \vecbf{E}_p^* \cdot \Jdm }
    {\int_{V_i} d^3x \; |\vecbf{E}_p|^2} \rp &.
}
Here we have taken $\Jdm$ to represent the magnitude of the oscillating harmonic DM source of frequency $\omega$, which is the same at each resonator location, and $\alpha_i$ is its phase which may vary between resonators.

The action of DM on this system of $N$ resonators is nicely characterized by thinking of the system collectively as one oscillator possessing $N$-cells and an $N$-fold degenerate subspace of normal modes with frequency $\omega_p$. 
The modes are labeled by the relative amplitude and phase of the field in each cell.  
Which of these normal modes does DM excite?  
It is to this mode that the readout must couple to draw the full available power. 

Suppose the DM coherence length is larger than the collective system, then the source phases are all equal $\alpha_i = \alpha_0$ and DM excites the uniform ``$0$-mode" in which all resonators are excited with equal amplitude and phase.  
In this case we ought to couple the readout to all resonators identically.  
However, now consider the opposite limit in which the DM coherence length is smaller than the smallest inter-resonator spacing. 
In this case DM still excites one particular normal mode, that of equal amplitudes but non-equal phases $\{e^{i \alpha_i}\}$. 
As the $\alpha_i$ are unknown, we do not a priori know to which mode we ought couple.  
Further, after a coherence time $1/(\mdm v^2) \sim \tx{ns} \, (\eV/\mdm)$ the relative phases $\alpha_i$ change by $\Order(1)$. 
\emph{In this case, the DM-driven normal mode is a moving target.}  
In order to extract all of the power, we must simultaneously readout a complete basis of $N$ normal modes.\footnote{
In principle this does not need to be simultaneous.
One could scan modes serially to identify the excited one and re-scan every coherence time.  
Such techniques are used in telecommunications, but it is unlikely they would be of use in a DM search where the signal is expected to be so weak that $\tx{SNR} \gtrsim 1$ is only achieved after integrating longer than the coherence time.}
A set of such modes acts as a net, catching all of the available DM power.

\subsection{Single-mode Readout}
An array of resonators coupled to a single output mode can only achieve complete coupling in the case of that DM is coherent over the full system. 
An example of such a readout is given in \cref{fig:series}a, in which one waveguide is coupled to $N$ identical resonators in series. 
In this case the bus couples to one normal mode of the $N$-cell system, that which has a relative phase between consecutive cells matching the optical length of waveguide between those cells. 
In a single-mode readout of $N$ incoherent resonators, the expected power is in fact independent of $N$ and is equal to the power delivered by a single resonator, as we show here.

Consider an arbitrary readout mode with electric field
\eqa{
\label{eq:random-mode}
    b \, \vecbf{E}_b\lp \vecbf{r}\rp = b \sum_i b_i \vecbf{E}_{p,i}\lp \vecbf{r}\rp,
}
where $\vecbf{E}_{p,i}$ is the mode function localized to resonator $i$ and $b$ is the overall mode amplitude.
This mode may be labeled by a complex $N$-tuple of the relative amplitudes, $\ket{b} = \lp b_1, b_2, \ldots , b_N \rp$.
How much power does this mode draw from the DM source? 
It obeys 
\eqa{
\label{eq:collective-eom}
    \lp \partial_t^2 + \frac{\omega_p}{Q} ( 1 + \gamma) \partial_t + \omega_p^2 \rp b \, \vecbf{E}_b 
    = - e^{i \alpha(r)} \partial_t \Jdm,
}
where we understand $\Jdm$ to be spatially uniform and $\alpha(r)$ is a spatially varying phase such that $\alpha(r) = \alpha_i$ is a constant inside the $i^\tx{th}$ resonator. 
Note that the quality factor of the collective system is equal to that of an individual resonator. 
The coupling of power out of this mode is parameterized by $\gamma$, which we define to be the ratio of signal power to dissipated power (equivalently, $1+\gamma$ is ratio between intrinsic and loaded quality factors of the \emph{collective} mode $b$ in \cref{eq:collective-eom}).  
Thus 
\eqa{
   \Psig = \frac{\omega \gamma}{2 Q} \int_\tx{N} d^3r \, |b \vecbf{E}_b|^2,
}
where the integral is over the volumes of all $N$ resonators. 
Since the modes are spatially disjoint and of equal amplitude, this is
\eqa{
   \Psig = \frac{\omega \gamma}{2 Q} 
   \ip{b}{b} \, |b|^2 \int_\tx{1} d^3r \, |\vecbf{E}_p|^2
}
and the integral is now only over the volume of one resonator. 
The amplitude $b$ follows from \Eq{eq:collective-eom}.
Specialising immediately to the on-resonance case we have  
\eqa{
    b &= \lp \frac{- i Q}{\omega_p \lp 1 + \gamma \rp} \rp
    \frac{\int_N d^3r \; \vecbf{E}_b^* \cdot \Jdm e^{i \alpha(r)}}
    {\int_\tx{N} d^3r \, |\vecbf{E}_b|^2} \\ 
     &= \lp \frac{- i Q}{\omega_p \lp 1 + \gamma \rp} \cdot 
     \frac{\int_1 d^3r \; \vecbf{E}_p^* \cdot \Jdm}
    {\int_\tx{1} d^3r \, |\vecbf{E}_p|^2}  \rp 
     \frac{\ip{b}{\alpha}}{\ip{b}{b}} .
\label{eq:collective-amplitude} 
}
Here we have associated the DM source with an $N$-tuple of phases, $\ket{\alpha} = \lp e^{i \alpha_1}, e^{i \alpha_2}, \ldots , e^{i \alpha_N} \rp$. 
Note that the factor in parenthesis in \Eq{eq:collective-amplitude} is the amplitude we would have in the case of a single resonator. 
The final signal power may be written 
\eqa{
   \Psig = N P_c \frac{4 \gamma}{\lp 1 + \gamma\rp^2} |\chi|^2
   \label{eq:NPc}
}
where $P_c$ is the maximal power that can be drawn from a single resonator, given in \Eq{eq:Pc}
and $|\chi|^2$ is an overlap factor between the readout mode $\ket{b}$ and the DM source mode $\ket{\alpha}$,
\eqa{
   |\chi|^2 = \frac{1}{N} \frac{| \ip{b}{\alpha} |^2}{\ip{b}{b}} .
}
The signal power is maximized for $\gamma = 1$. 
The overlap $\chi$ is proportional to the projection of the DM mode onto the readout mode, and as such its maximal value occurs when these vectors are parallel,
\eqa{
    |\chi|^2 \leq \frac{\ip{\alpha}{\alpha}}{N}  = 1
}
where $\ip{\alpha}{\alpha} = N$ since it is a tuple of phases.  

In the case of DM coherent over all resonators, all $\alpha_i$ are equal and we may choose a single readout mode in which all $b_i$ are equal, for example by spacing the resonators at integer multiples of the wavelength along the bus. 
It follows then that $|\chi|^2 = 1$ and the output power is optimal, $\Psig = N P_c$.
However, in the fully incoherent case, for a single readout  mode we ought to take the $b_i$ fixed and average over the $\alpha_i$, which yields a suppressed overlap,
\eqa{
    \langle | \chi |^2 \rangle_\alpha = \frac{1}{N}
}
and thus $\Psig = P_c$ for any fixed readout mode $\{ b_i \}$.

\subsection{Multi-mode Readout}
\label{sec:multimode}

In the incoherent case, the full power is recovered by reading out a full basis of normal modes.  
For any possible DM excitation mode, its overlap with this set of readout modes will be sufficient to extract all of the power.  
To demonstrate this, suppose that one reads out an orthogonal set of modes $\ket{c}$ labeled by $c = 0, 1, \, \ldots \, N-1$, and each of these is critically coupled ($\gamma = 1$) to the output.  
Then the total output power is given by the sum of overlap factors, 
\eqa{
    \sum_c \Psig = N P_c \sum_c |\chi \lp c \rp |^2,
}
and the sum of overlaps is $1$, since it is a sum of projections onto an orthogonal basis, i.e., 
\eqa{
    &\sum_c \frac{\ket{c} \bra{c}}{\ip{c}{c}} = 1
}
and the sum of overlaps is 
\eqa{
    \sum_c |\chi|^2 = \frac{1}{N} &\sum_c \frac{\ip{\alpha}{c} \ip{c}{\alpha}}{\ip{c}{c}} = 1
}
which follows since $\ip{\alpha}{\alpha} = N$. 
Thus the summed power is optimal, $\sum_c \Psig = N P_c$.

\section{Numerical Studies using the Coupled Mode Model}
\label{sec:cmt}

In order to develop intuition for the output power in \cref{eq:fourier_power_amp_squared,eq:dm_source_ensemble_average} we will apply them to $N=1$ and $N=2$ before presenting numerical results for large $N$.

\subsection{One Resonator}\label{sec:one_resonator}
Evaluating \crefrange{eq:sig_power_in_one_direction}{eq:dm_source_ensemble_average} for $N=1$ gives the total output power of a single resonator 
\beq
P_1(m) = \frac{\tau^{-1}_{\mathrm{e}} \bar{J}^2_{\mathrm{DM}} |\eta|^2 V}{\omega_R^2}\frac{m^2}{(m-\omega_R)^2 + \tau^{-2}} 
\eeq
where we used $\tau^{-1}_{\mathrm{e}} = |\kappa|^2$ and added power radiated into both directions in the bus. On resonance, $m \approx \omega_R$, this becomes
\beq
P_{1,\mathrm{res}}(m) =\frac{2Q}{m}\left(\frac{\tau_{\mathrm{e}}^{-1}}{\tau^{-1}}\right)\bar{J}_{\mathrm{DM}}^2 V |\eta|^2,
\label{eq:P1_res}
\eeq
where $Q$ is the loaded quality factor and $\tau^{-1}_{\mathrm{e}}/\tau^{-1}$ encodes the ``branching fraction'' of the resonator to decay into the bus. 
We define the ratio $\beta = \tau^{-1}_{\mathrm{e}}/\tau_\mathrm{i}^{-1}$ so that 
\eqa{
\label{eq:def-beta}
    \frac{\tau_{\mathrm{e}}^{-1}}{\tau^{-1}} = \frac{Q}{Q_\mathrm{e}} = \frac{\beta}{1 + \beta}
}
Because $\tau^{-1} = \tau_{\mathrm{i}}^{-1} + \tau_{\mathrm{e}}^{-1}$, the signal power is maximized when $\tau_{\mathrm{i}}^{-1} = \tau_{\mathrm{e}}^{-1}$, i.e.~$\beta = 1$. 
For this choice, \cref{eq:P1_res} agrees with~\cref{eq:generic_signal_power}.

We will refer to \emph{critical coupling} as any choice of $\kappa_l$ that maximizes the power output of a given system.  
In general, the critical coupling depends on the number of resonator on the bus and their frequencies; moreover each resonator can have a different coupling, leading to a high-dimensional optimization problem for $\kappa_l$. For simplicity, we will focus on the case where all of these couplings are the same.

\subsection{Two Resonators}
\begin{figure}[t]
     \centering
  \includegraphics[width=8cm]{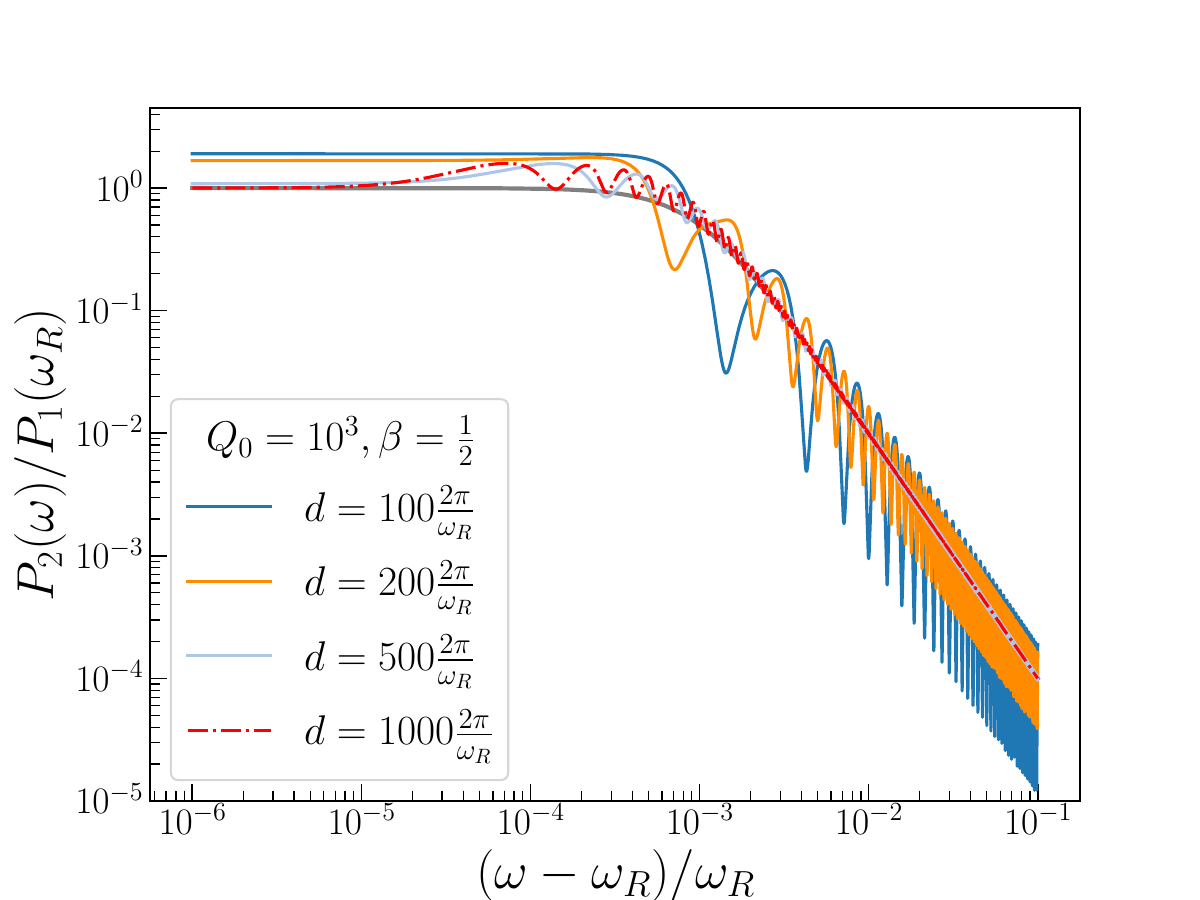}  
\caption{Signal power from two resonators for different separations $d$ as a function of the relative DM frequency detuning. For comparison, the single resonator power (gray) with $\beta=1$ is plotted as well. $v_0=10^{-3},n_r=3.5$}
\label{fig:two_rings}
\end{figure}

When $N=2$ we can write down a short, closed form expression for the signal power as a function of resonator spacing, read out couplings and 
DM frequency detuning. Letting $\omega=\omega_R(1+\Delta)$, the power is given by
\begin{widetext}
\beq
\begin{split}
&P_{{\rm sig},\; 2}(\Delta,\beta)=
\frac{Q_{\rm i}  \bar{J}_{\mathrm{DM}}^2  |\eta|^2 V}{\omega_R}\beta\times\\
&\left[\frac{ -\beta  \rho  \cos (2 \alpha +\theta )+\beta ^2+\beta +4 \Delta ^2 Q_{\rm i}^2+1}{ -2 \beta ^2 \rho ^2 \cos (2 (\alpha +\theta ))+\beta ^4+\rho ^4}
+e^{-\frac{1}{4} (\Delta +1)^2 v_0^2 \omega_R^2d^2} \frac{(1+4 \Delta ^2 Q_{\rm i}^2) \cos \alpha +4 \beta  \Delta  Q_{\rm i} \sin \alpha }{ -2 \beta ^2 \rho ^2 \cos (2 (\alpha +\theta ))+\beta ^4+\rho ^4}\right]
\end{split}
\eeq
\end{widetext}
where
\begin{subequations}
\begin{align}
\alpha(\Delta) & =(\Delta +1) n_r \omega_Rd\\
\rho(\Delta,\beta)& =\sqrt{(\beta +1)^2+4 \Delta ^2 Q_{\rm i}^2}\\
\theta(\Delta,\beta)&=\tan ^{-1}\left(\frac{2 \Delta Q_{\rm i}}{\beta +1}\right)
\end{align}
\end{subequations}
In \cref{fig:two_rings} we compare the power output from two critically coupled resonators for different separations $d$. It is clear that as the separation becomes comparable to the DM coherence length $\sim\frac{v_0^{-1}}{\pi}\frac{2\pi}{\omega_R}$, the resonance power deteriorates, and increasingly approaches the single resonator power as the separation gets further.

\subsection{$N$ Resonators, Coherent Sources}
\label{sec:coherent_same_freq}
Next, let us consider the response when $\omega=\omega_R$ and the DM sources are all coherent, i.e.,  $s_{\mathrm{DM},i}$ all have the same phase. This means that the resonator spacing $d$ must be small enough such that $v_0\omega_R d N\ll1$. Furthermore, let us choose $d$ such that $\omega_R n_r d$ is an integer multiple of $2\pi$. One can show that the contribution to $s_{\rm out}$ from each resonator takes the following form
\beq
 \frac{\beta^{1/2}}{\beta+N^{-1}}\left(\frac1N,\frac1N,\frac1N,\cdots\right),
\eeq
where $\beta$ is defined by $Q = Q_{\rm i}/(1+\beta)$. 
The total signal power then scales as
\beq\label{eq:N_res}
P_{\mathrm{sig},\; N,\;{\rm coherent}}(\omega=\omega_R)=\frac{Q_{\rm i}}{\omega_R}\frac{N^2\beta}{(1+N\beta)^2}\bar{J}_{\mathrm{DM}}^2  |\eta|^2 V
\eeq
The coupling that maximizes the output is $\beta=1/N$. This corresponds to a ``critical coupling'' for the entire system for which the signal power becomes:
\beq
P_{\mathrm{sig},\; N,\;{\rm coherent},\;{\rm crit}}(\omega=\omega_R)=\frac{NQ_{\rm i}}{\omega_R}\bar{J}_{\mathrm{DM}}^2  |\eta|^2 V.
\eeq
Compared with \cref{eq:P1_res} evaluated at critical coupling (which includes an additional factor for $2$ for radiation into both directions in the bus), it is clear that the signal power grows linearly with the number of resonators as long as they are within the coherence length of the dark matter.

Using $Q_{\rm i}=\beta Q_{\rm e}$, followed by taking $\beta\to \infty$, \cref{eq:N_res} becomes
\beq
P_{\mathrm{sig},\;N\;{\rm coherent},\;\text{no loss}}(\omega=\omega_R)=\frac{1}{2}\tau_{\rm e } \bar{J}_{\mathrm{DM}}^2  |\eta|^2 V.
\eeq
This is the limit where $\tau_{\mathrm{i}} \ll \tau_{\mathrm{e}}$, i.e., internal losses are unimportant. In this regime, the output power on resonance is \emph{exactly} that of a single resonator. This has a simple intuitive interpretation in the fact that signal photons are trapped in the ``bulk'' of the array by being repeatedly re-absorbed by the 
different resonators; as a result the only photons that are detected come from resonators on either end of the array. 

\subsection{$N$ Resonators, Incoherent Sources}
\label{sec:N_resonators}
While the system in \cref{eq:cmt_equation} can be solved analytically for a few resonators, the general formulas are 
not enlightening, even for $N=2$. Instead we study the solutions numerically and discuss special limiting cases. Our first goal is to investigate the $N$ scaling of the signal power. As we will see it crucially depends on whether the 
resonators have equal resonant frequencies or not. 

\begin{figure}
\centering
\includegraphics[width=8cm]{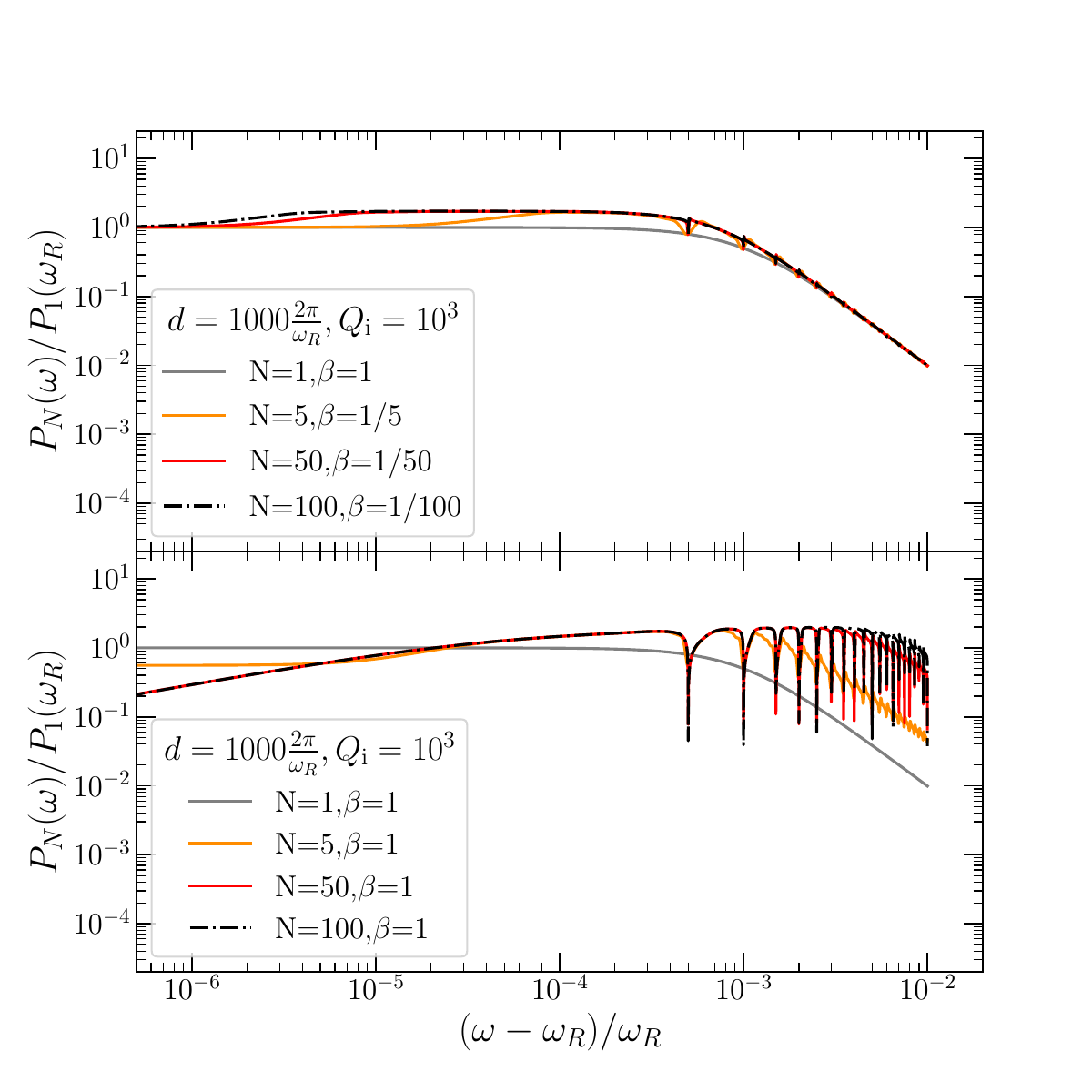}
\caption{The power drawn from $N$ resonators with the \emph{same frequency} on one bus (\cref{eq:sig_power_in_one_direction}), normalized by $P_{\rm res,1}$, the peak power from a single critically-coupled resonator (\cref{eq:P1_res}).
$\beta$ is the coupling ratio defined in \Eq{eq:def-beta}, taken to be 1 in the bottom panel, and critical for $N$ coherently spaced resonators in the top panel, however these have spacing $d$ which is beyond coherence. 
}\label{fig:power_samefrequency}
\end{figure}

First, let us suppose that the $\omega_{R_l} = \omega_R$ for all $l$. 
In \cref{fig:power_samefrequency} we show the signal power as a function of relative detuning $(\omega_R - m)/\omega_R$ for different choices of $N$. The main conclusion that we draw from this figure is that peak power does not scale with $N$ for generic choices of the resonator-bus couplings. The physical reason for this is that a signal photon produced by one resonator travelling down the bus can resonantly excite any of the other resonators, which provides more opportunity for the photon to be lost. Specifically, the probability of a bus photon to resonantly excite a resonator and then to be lost due to internal losses is $\sim N \tau^2 / (\tau_{\mathrm{i}}\tau_{\mathrm{e}})$ (unless it originates close to the end of the array, in which case, the $N$ enhancement is absent). 
If $\tau_{\mathrm{i}}\sim\tau_{\mathrm{e}}$ then in this resonant regime $N\sim 1$ already leads to an $\mathcal{O}(1)$ probability for losing the signal photon! Naively, there are two ways out of this predicament. For example, if $\tau_{\mathrm{e}} \gg \tau_{\mathrm{i}}$ (and so $\tau \approx \tau_{\mathrm{i}})$, the loss probability is $N \tau_{\mathrm{i}} / \tau_{\mathrm{e}}$; so by varying $\tau_{\mathrm{e}}$ via the bus resonator couplings one can try to keep this probability small as $N$ becomes large. This has two challenges: first, each resonator coupling needs to be adjusted based on its position in the array, leading to a high-dimensional optimization problem; second, taking $\tau_{\mathrm{i}} / \tau_{\mathrm{e}}$ small also shrinks the output power of each resonator (c.f., \cref{eq:P1_res}) as $1/N$, so the overall $N$ scaling is still lost, as we see in the $(\omega_R - m)/\omega_R \to 0$ corner of \cref{fig:power_samefrequency}.

Both arguments about the lack of $N$ scaling assumed that $m = \omega_R$; away from this resonant regime, the output power \emph{does} grow with $N$, but the $Q$ enhancement is lost. \footnote{$N$ scaling is also present if all $N$ resonators are within a single DM coherence length; this, however, means that each resonator has a volume that is a factor $\sim 1/N$ smaller than it could be.} 
Thus there appears to be no advantage of having resonators of the same frequency coupled to a single bus with fixed phase, as in the model 
of \cref{eq:cmt_equation,eq:output_power_amplitude_one_direction}.
In this setup only a single normal mode of the entire system is read out. As discussed in \cref{ap:SI-combining} a multimode read out can restore the extensive $N$ scaling. 

\begin{figure}
\centering
\includegraphics[width=8cm]{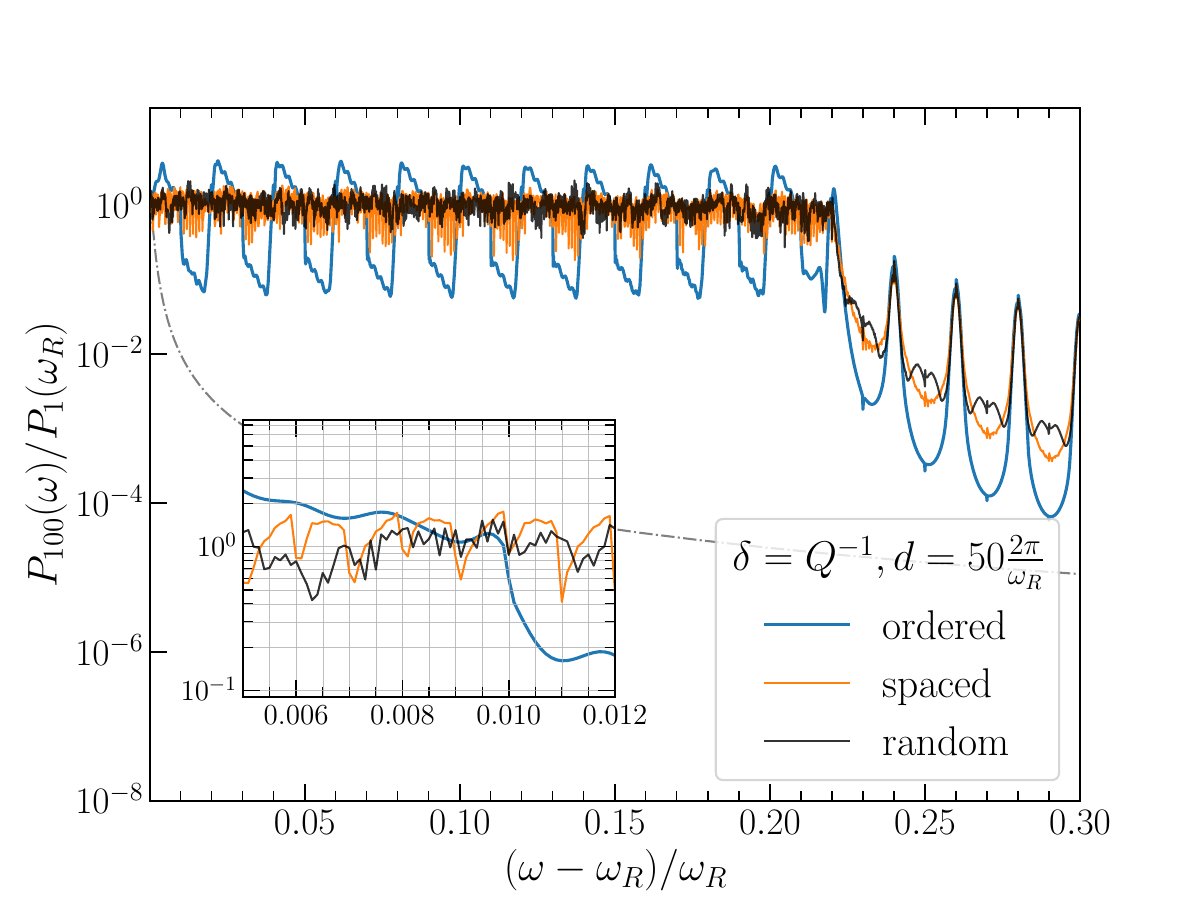}
\caption{Power output from $N=100$ resonators in series with three different spacial arrangements of resonant frequencies from the set in \cref{eq:omega_i}. Each resonator has $Q=500$. $P_1(\omega_R)$ is the peak power from a single resonator. The total signal bandwidth is given by $\sim N \omega_R/ Q$. For comparison, $P_1(\omega)/P_1(\omega_R)$ is shown in a gray dashed line.
The spacing between resonators is less than the coherence length, which gives rise to the clear structure in the ordered case. 
When the resonators are randomized or spaced according to \Eq{eq:spaced} this structure disappears as all resonators with similar frequencies are then incoherent. 
\label{fig:width}
}
\end{figure}

We now consider the case where all the resonance frequencies are different, drawn from 
the set
\beq\label{eq:omega_i}
\omega_{R_l} \in \{ \omega_R(1+\delta )^{i-1} \;\; | \;\; i \in [1..N]\}
\eeq
where $\omega_R$ is the lowest target frequency in the series and  $\delta$ is a fractional frequency change between neighbouring frequencies (these frequency space neighbours do not need to be neighbours in position space).
Since we are interested in covering a region of DM mass range without gaps, we will consider 
$\delta \sim \mathrm{few} \times Q^{-1}$. Because no two resonators have the same resonant frequency, 
the response of the coupled system is approximately the sum of responses of $N$ uncoupled resonators. 
This can be easily seen from the structure of \crefrange{eq:M}{eq:fourier_power_amp_squared}: at a given driving frequency $\omega$ only one of the diagonal entries in \cref{eq:M} is large, so the output power approximately decomposes into a sum over different resonators. This decomposition is nearly exact when the resonators are spaced by more than a de Broglie wavelength; otherwise non-trivial source phase correlation can play a role in determining the precise response, especially if the neighbouring resonators in physical space also have similar frequencies, as illustrated in \cref{fig:width}.
It is clear that the signal bandwidth grows with $N$ as $N \delta \omega_R$, enabling a broadband DM search in all three cases. The ordered arrangement corresponds to the resonators arranged sequentially according to frequency, i.e., $\omega_{R_l}=\omega_{R}(1+\delta)^{l-1}$. For the spaced arrangement neighbouring frequencies are separated in space by $10$ positions; this is implemented by taking $\omega_{R_l} = \omega_R(1+\delta)^{g(l)-1}$, where 
\beq\label{eq:spaced}
g(l)=
\begin{cases}
   \left\lfloor \frac{l}{10}\right\rfloor +\frac N{10} ((l \bmod 10)-1)+1  & l \bmod 10\neq 0 \\
   (l+9 N)/10  & l \bmod 10=0
\end{cases}
\eeq
assuming $N$ is a multiple of 10. For example, the first resonator has $\omega_{R_1} = \omega_R$, while the next frequency appears at $l=11$, i.e., $\omega_{R_{11}} = \omega_R(1 + \delta)$. Finally, the random arrangement allocates frequencies from \eqref{eq:omega_i} to the resonators randomly. 
It can be seen that the power becomes more evenly distributed among all frequencies within the bandwidth using a random or spaced arrangement compared to an ordered arrangement. This is because a small separation in frequencies between nearby resonators enables interference effects if they are within the DM coherence length. 
Therefore, the random or variations of the spaced arrangement of frequencies are preferred to ensure a ``smooth'' detector response.
This may arise naturally due to fabrication imprecision, or could be tuned with thermo-optics or local material deposition. 

For the randomly spaced array the peak power for a DM mass within the bandwidth of the resonator array is simply given by ~\cref{eq:P1_res}.

\onecolumngrid
\bibliography{biblio}


\end{document}